\documentclass[showpacs,reprint,amsmath,amssymb,aps,prb,floatfix,twocolumn,superscriptaddress]{revtex4}

\usepackage{graphicx}% include figure files
\usepackage{dcolumn}% align table columns on decimal point
\usepackage{bm}% bold math
\usepackage{verbatim}
\usepackage{color}

\newcommand{\dd}{\mathrm{\mathrm{d}}} 
\newcommand{\ket}[1]{\left| #1 \right\rangle}

\begin{document}

\title{Double-expansion impurity solver for multiorbital models with dynamically screened $U$ and $J$}
\author{Karim Steiner}
\affiliation{Department of Physics, University of Fribourg, 1700 Fribourg, Switzerland}
\author{Yusuke Nomura}
\affiliation{Centre de Physique Th\'eorique (CPHT), \'Ecole Polytechnique, 91128 Palaiseau Cedex, France}
\author{Philipp Werner}
\affiliation{Department of Physics, University of Fribourg, 1700 Fribourg, Switzerland}

\date{\today}

\begin{abstract}
We present a continuous-time Monte Carlo impurity solver for multiorbital impurity models which combines a strong-coupling hybridization expansion and a weak-coupling expansion in the Hund's coupling parameter $J$. This double-expansion approach allows to treat the dominant density-density interactions $U$ within the efficient segment representation. We test the approach for a two-orbital model with static interactions, and then explain how the double-expansion allows to simulate models with frequency dependent $U(\omega)$ and $J(\omega)$. The method is used to investigate spin state transitions in a toy model for fullerides, with repulsive bare $J$ but attractive screened $J$.  
\end{abstract}


\maketitle

\section{Introduction} 

Strongly correlated materials exhibit a range of interesting properties, such as unusually large susceptibilities or high-temperature superconductivity. A theoretical investigation of this class of materials is possible within the framework of dynamical mean-field theory (DMFT),\cite{Georges_1996} either at the simple model level or in combination with input from density-functional based {\it ab initio} calculations.\cite{Kotliar_2008} Because of the interest in cuprate high-temperature superconductors, and also because of algorithmic limitations, much effort has in the past been devoted to the study of the single-band Hubbard model. The discovery of aromatic superconductors,\cite{Mitsuhashi2010} iron pnictides\cite{Kamihara2006} and recent studies emphasizing the multi-band character of cuprates\cite{Sakakibara_2012} have however shifted some of the attention to correlated multi-orbital systems. In these multi-orbital systems, the Hund's coupling parameter plays an essential role and leads to new types of correlation phenomena, such as bad-metal behavior due to local-moment formation,\cite{Werner2008, Haule2009, Georges2013} magnetism and orbital ordering,\cite{Karsten_ferro,PhysRevB.58.R567,PhysRevLett.99.216402,PhysRevB.80.235114} spin-state transitions,\cite{Werner2007_crystal, Kunes2009} orbital-selective Mott transition,~\cite{PhysRevLett.102.126401,PhysRevB.83.205112} nontrivial spatial correlations,\cite{PhysRevB.79.245128,nomura_spatial} and unconventional superconductivity.\cite{Hoshino2015} 

Dynamical mean field simulations of generic multi-band models have become possible thanks to the development of strong-coupling (hybridization expansion) impurity solvers.\cite{Werner_2006} The matrix\cite{Werner_2006_Matrix} or Krylov-implementations\cite{Laeuchli_2009} of this impurity solver can handle arbitrary interactions among the orbitals, but the computational effort scales exponentially with the number of orbitals. A far more efficient simulation is possible within the so-called segment formalism,\cite{Werner_2006} if the interactions are restricted to the density-density component of the full Coulomb matrix, and this approximation is still often made in simulations of  transition metal and actinide compounds. In most cases these density-density terms give the dominant contribution to the interaction energy, so that it may be advantageous to consider the spin-flip, pair-hopping, and correlated hopping terms as a perturbation in an expansion that treats the density-density components exactly. In this paper, we explore such a double-expansion impurity solver, which stochastically samples a diagrammatic expansion of the impurity partition function in powers of the hybridization function and the interaction terms which are not of density-density type. Such an impurity solver trades the exponential scaling of the matrix/Krylov approach with an additional weak-coupling type expansion, and (for more than two orbitals) a potential sign problem. It should be efficient in the case of a small number of orbitals and not too large Hund's coupling. Here, we implement and test the double-expansion solver for a two-orbital model with rotationally invariant interaction.

Besides potential efficiency gains, a second important reason for exploring the double-expansion approach is that such a solver enables the simulation of certain types of problems which cannot be solved using the established hybridization expansion methods. A relevant example is models with a dynamically screened $J$. The low-energy effective models solved in DMFT simulations of correlated materials can be obtained from a down-folding procedure in which the bands outside some energy window around the Fermi level are integrated out.\cite{Aryasetiawan_2004} This procedure leads to a dynamically screened interaction. For example, in transition metals and their compounds, the screening of charge fluctuations results in density-density interactions which range from a bare value of typically about 20 eV to a screened value of only a few eV. Highly efficient algorithms exist to treat this type of screening.\cite{Werner_2007, Werner_2010} The screening of the Hund's coupling parameter is usually much weaker, so that the bare and screened $J$ typically differ by less than 20\%.\cite{Sasioglu2011} In all the DMFT simulations to date, the Hund's coupling has thus been treated as frequency-independent. However, given the sensitivity of multi-orbital phase diagrams on the Hund's coupling parameter\cite{Werner2008, Werner2009,Medici2011} it is desirable to develop a method which extends the efficient technique of Ref.~\onlinecite{Werner_2010} to models with a frequency-dependent $J(\omega)$. There are also materials in which the dynamical screening of $J$ plays a crucial role. In alkali-doped fullerenes,\cite{C60_super1,RevModPhys.69.575} the observed superconductivity is believed to arise from an overscreened $J$:\cite{Capone_2002,RevModPhys.81.943,nomura_C60_paper} as a result of Jahn-Teller screening $J(\omega)$ turns negative at some low frequency and hence favors low-spin states. The double-expansion solver allows simulations of such dynamically screened multi-orbital systems.

The structure of the paper is as follows. In Section~\ref{sec:method} we describe the double expansion method for the case of a two-orbital model with rotationally invariant interactions and explain how this method can be used to treat models with dynamically screened $U$ and $J$. Section~\ref{sec:tests} shows some test results and information on the average perturbation orders. Simulation results for a model with static $U$ and dynamical $J(\omega)$ are presented in Section~\ref{sec:results}, and a brief summary and outlook is given in Section~\ref{sec:conclusions}.

\section{Model and Method}\label{sec:method}

\subsection{Two-orbital model}

As a simple, but nontrivial example, we consider a two-orbital model with rotationally invariant Slater-Kanamori interactions and a possible crystal field splitting. DMFT replaces the lattice problem by the self-consistent solution of a two-orbital quantum impurity model with Hamiltonian 
\begin{eqnarray} \label{eq:hamil}
	\mathcal{H} &=& \mathcal{H}_{\mathrm{dens}} 
	+ \mathcal{H}_{\mathrm{sf}}  + \mathcal{H}_{\mathrm{sf}}^{\dagger}
	+ \mathcal{H}_{\mathrm{ph}}  + \mathcal{H}_{\mathrm{ph}}^{\dagger}\nonumber\\
	&&+ \mathcal{H}_{\mathrm{bath}}  
	+ \mathcal{H}_{\mathrm{hyb}} + \mathcal{H}_{\mathrm{hyb}}^{\dagger}.
\end{eqnarray}
The density-density, spin-flip, pair-hopping, bath and hybridization parts of the Hamiltonian are given by
\begin{eqnarray}
	\mathcal{H}_{\mathrm{dens}} &=& -\sum_{\alpha,\sigma} \mu n_{\alpha, \sigma} 
		+ \sum_{\sigma} \Delta ( n_{1, \sigma} - n_{2, \sigma} ) \nonumber\\
		&&+ \sum_{\alpha} U n_{\alpha, \uparrow} n_{\alpha, \downarrow} + \sum_{\sigma} U' n_{1, \sigma} n_{2, -\sigma} \nonumber\\
		&& + \sum_{\sigma} ( U' - J ) n_{1, \sigma} n_{2, \sigma},\hspace{4mm} \\
	\mathcal{H}_{\mathrm{sf}} &=& - J c_{1, \downarrow}^{\dagger} c_{2, \uparrow}^{\dagger} 
		c_{2, \downarrow} c_{1, \uparrow}, \\
	\mathcal{H}_{\mathrm{ph}} &=& - J c_{2, \uparrow}^{\dagger} c_{2, \downarrow}^{\dagger}
		c_{1, \uparrow} c_{1, \downarrow}, \\
		\mathcal{H}_{\mathrm{bath}} &=& \sum_{k, \alpha,\sigma} \epsilon_{k} a_{k,\alpha,\sigma}^{\dagger} a_{k,\alpha,\sigma}, \\ 
	\mathcal{H}_{\mathrm{hyb}} &=& \sum_{k, \alpha,\sigma} c^\dagger_{\alpha,\sigma} V_{k,\alpha,\sigma} a_{k,\alpha,\sigma}, 
\end{eqnarray}
where $\alpha=1,2$ is the orbital index, $\sigma=\uparrow,\downarrow$ (or $\pm 1$) the spin index, $\Delta$ the crystal field splitting, $U$ the intra-orbital interaction, and $J$ the coefficient of the Hund coupling. We choose the interorbital interaction $U'=U-2J$ for rotational invariance and denote the impurity creation operators by $c^\dagger_{\alpha,\sigma}$ and the density operators by $n_{\alpha,\sigma}=c^\dagger_{\alpha,\sigma} c_{\alpha,\sigma}$. The bath levels, with creation operators $a^\dagger_{k,\alpha,\sigma}$ and energy $\epsilon_k$ are parametrized by a quantum number $k$. The bath energies and the hybridization parameters $V_{k,\alpha,\sigma}$ define the hybridization function
\begin{equation}
\Lambda_{\alpha,\sigma}(i\omega_n)=\sum_k\frac{|V_{k,\alpha,\sigma}|^2}{i\omega_n-\epsilon_k}.
\end{equation} 
In the case of a semi-circular density of states with bandwidth $4t_\alpha$, the DMFT self-consistency condition provides a simple relation between the hybridization functions and impurity Green's functions $G_{\alpha,\sigma}$:\cite{Georges_1996}
\begin{equation}
\Lambda_{\alpha,\sigma}=t^2_\alpha G_{\alpha,\sigma}.
\end{equation}
We will consider an orbital-independent semi-circular density of states with $t_\alpha=t$ ($\alpha=1,2$) and use $t$ as the unit of energy.

\subsection{Double expansion for static interactions}
\label{sec_static}

We solve the impurity model (\ref{eq:hamil}) using the continuous-time Monte Carlo technique.~\cite{Gull2011} In the double-expansion approach, this continuous-time method is based on a simultaneous expansion of the partition function in the hybridization terms and the interaction terms which are not of density-density type (in the model considered here, the spin flip and pair hopping terms). To derive the formalism, we switch to an interaction representation in which the time-evolution of operators is given by $\mathcal H_\text{dens}+\mathcal H_\text{bath}$ and write the partition function of the impurity model as
\begin{align} \label{eq:partfunc}
		&Z =\mathrm{Tr}_{c}  \mathrm{Tr}_{b} \left[ e^{-\beta(\mathcal H_\text{dens}+\mathcal H_\text{bath})}T_{\tau} \exp \left( 
		- \int_0^{\beta} \dd \tau   \big[
		\mathcal{H}_{\mathrm{sf}}(\tau)\right. \right.\nonumber\\
		&  + \left. \left. \vphantom{\int_0^{\beta}}    \mathcal{H}_{\mathrm{sf}}^{\dagger}(\tau) 
		+ \mathcal{H}_{\mathrm{ph}}(\tau)  + \mathcal{H}_{\mathrm{ph}}^{\dagger}(\tau) 
		+\mathcal{H}_{\mathrm{hyb}}(\tau) + \mathcal{H}_{\mathrm{hyb}}^{\dagger}(\tau) \big]
	\right)  \right].
\end{align}
The next step is to expand the time-ordered exponential in powers of the spin-flip, pair-hopping and hybridization terms. Since there is then no coupling between the impurity and the bath anymore in the time-evolution (given by $\mathcal H_\text{dens}+\mathcal H_\text{bath}$), the trace over the bath states can be computed analytically.\cite{Werner_2006, Werner_2006_Matrix} This leads to the expression
\begin{align}
&\frac{Z}{Z_\text{bath}}=\sum_{\{n_{\alpha,\sigma}\}}\sum_{n_\text{sf}}\sum_{n_\text{ph}}
\left(\prod_{\alpha,\sigma}\int_{\tau_{h_1}<\ldots<\tau_{h_{n_{\alpha,\sigma}}}}\int_{\tau'_{h_1}<\ldots<\tau'_{h_{n_{\alpha,\sigma}}}}\right)\nonumber\\
&\times\int_{\tau_{s_1}<\ldots<\tau_{s_{n_\text{sf}}}}\int_{\tau'_{s_1}<\ldots<\tau'_{s_{n_\text{sf}}}}\nonumber\\
&\times\int_{\tau_{p_1}<\ldots<\tau_{p_{n_\text{ph}}}}\int_{\tau'_{p_1}<\ldots<\tau'_{p_{n_\text{ph}}}} 
w(\tau_{h_1},\ldots,\tau'_{n_{\text{ph}}}),
\end{align}
where $Z_{\mathrm{bath}} = \mathrm{Tr}_b e^{-\beta \mathcal{H}_{\mathrm{bath}}}$ and the weight of a configuration consisting of $2n_h$ hybridization 
events ($n_h = \sum_{\alpha, \sigma} n_{\alpha, \sigma}$), $2n_\text{sf}$ spin-flip events and $2n_\text{ph}$ pair hopping events is given by  
\begin{align} \label{eq:weight}
	w &= \mathrm{Tr}_{c} \Bigg[ e^{-\beta 
		\mathcal{H}_{\mathrm{dens}}} T_{\tau} \prod_{\alpha,\sigma}  
		c_{\alpha,\sigma} ( \tau_{h_{n_{\alpha,\sigma}}} ) c^{\dagger}_{\alpha,\sigma} ( \tau_{h_{n_{\alpha,\sigma}}}' ) \dots\nonumber\\ 
		& \hspace{50mm}c_{\alpha,\sigma} ( \tau_{h_1} ) c^{\dagger}_{\alpha,\sigma} ( \tau_{h_1}' ) \nonumber\\
		& \times S( \tau_{s_{n_\text{sf}}} ) S^\dagger( \tau'_{s_{n_\text{sf}}} )
		\dots S( \tau_{s_1} ) S^{\dagger} ( \tau_{s_1}' ) \nonumber\\
		& \times P( \tau_{p_{n_\text{ph}}} ) P^{\dagger} ( \tau'_{p_{n_\text{ph}}} )
		\dots P( \tau_{p_1} ) P^{\dagger} ( \tau_{p_1}' ) \Bigg] \nonumber\\ 
		& \times \prod_{\alpha,\sigma} \mathrm{det} (M_{\alpha,\sigma}^{-1} (\{\tau_h\},\{\tau_h'\} ) ) \nonumber\\
		& \times J^{2n_{\mathrm{sf}}} J^{2n_{\mathrm{ph}}}(\dd\tau)^{2(n_h+n_\mathrm{sf}+n_\mathrm{ph})},
\end{align}
where the time evolution of operators is now given by $\mathcal H_\text{dens}$, $S = -c_{1, \downarrow}^{\dagger} c_{2, \uparrow}^{\dagger} c_{2, \downarrow} c_{1, \uparrow}$, 
$P = -c_{2, \uparrow}^{\dagger} c_{2, \downarrow}^{\dagger}c_{1, \uparrow} c_{1, \downarrow}$, and $M_{\alpha,\sigma}^{-1}$ is a $n_{\alpha,\sigma}\times n_{\alpha,\sigma}$ matrix of hybridization functions, with elements $M_{\alpha,\sigma}^{-1}(i,j) = \Lambda_{\alpha,\sigma}(\tau'_{h_i} - \tau_{h_j})$.

\begin{figure}
	\includegraphics[width=9cm]{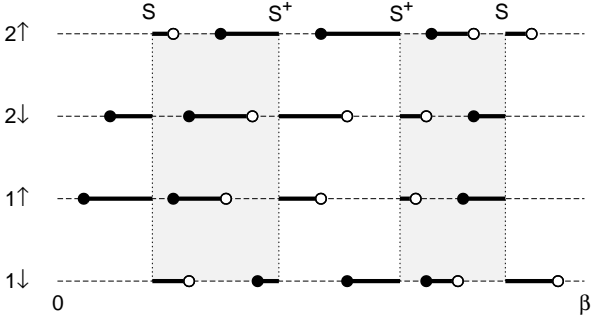}
	\caption{\label{fig_sf} 
	Illustration of an order $n_h = 12$, $n_{\textrm{sf}} = 2$ configuration for the 
	two orbital model. Empty (full) circles represent annihilation (creation) operators. 
	Vertical dashed lines indicate spin-flip events.
}
\end{figure}

\begin{figure}
	\includegraphics{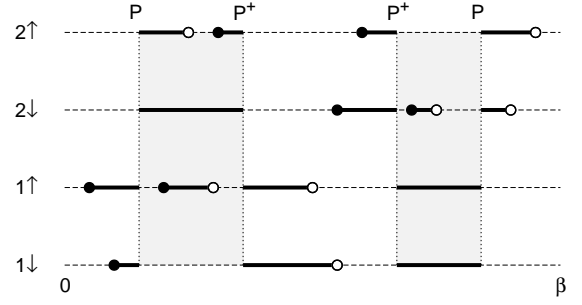}
	\caption{\label{fig_ph} 
	Illustration of an order $n_h = 7$, $n_{\textrm{ph}} = 2$ configuration for the 
	two orbital model. Empty (full) circles represent annihilation (creation) operators. 
	Vertical dashed lines indicate pair-hopping events.
}
\end{figure}

The trace vanishes unless there is an equal number of impurity creation and annihilation
operators for each flavor.  This requirement implies that for each $\mathcal{H}_{\mathrm{hyb}}$, we must have a corresponding 
$\mathcal{H}_{\mathrm{hyb}}^{\dagger}$, and similarly for the spin-flip and pair-hopping terms. The expansion in the spin-flip and pair-hopping
terms hence does not lead to a sign problem in this two-orbital case with static interactions. Furthermore, because the time-evolution operator is diagonal in the occupation number basis, 
we can use the segment representation\cite{Werner_2006} to graphically represent all the non-vanishing contributions to the trace (see Figs.~\ref{fig_sf} and \ref{fig_ph}).
In this representation, each segment marks a time-interval in which the impurity is occupied by an electron of a given flavor ($\alpha,\sigma$). We denote such a segment configuration by $C$ and sample the space of all configurations using the Metropolis algorithm.    

\subsection{Monte Carlo sampling}
\label{MCsampling}

The expression for the trace in Eq.~(\ref{eq:weight}) shows that a given configuration $C$ consists of $2n_h$ hybridization events, $2n_\text{sf}$ spin-flip events, and $2n_\text{ph}$ pair-hopping events. We can generate all possible configurations using local updates which insert or remove pairs of hybridization, spin-flip or pair-hopping events. These updates must satisfy the detailed-balance condition $w( C ) p(C\rightarrow C')=w(C')p(C'\rightarrow C)$, where $p( C\rightarrow C')$ is the transition probability from configuration $C$ to configuration $C'$. We split this transition probability into a proposal probability and an acceptance probability, $p(C\rightarrow C')=p^\text{prop}(C\rightarrow C')p^\text{acc}(C\rightarrow C')$ and define the ratio of acceptance probabilities 
\begin{equation}
R(C\rightarrow C') \equiv \frac{p^\text{acc}(C\rightarrow C')}{p^\text{acc}(C'\rightarrow C)}=\frac{p^\text{prop}(C'\rightarrow C)}{p^\text{prop}(C\rightarrow C')}\frac{w(C')}{w( C )}.
\end{equation}
In the Metropolis scheme, the update is accepted with probability $\text{min}[1,R]$.

For the hybridization events, the sampling procedure is exactly the same as detailed in Ref.~\onlinecite{Werner_2006}. For the insertion, we try to place a creation operator at a randomly chosen time $\tau'_h$ on the imaginary-time interval. If it falls on a segment, the move is rejected. Otherwise, we compute the length $l_{\max}$ of the interval to the next creation operator (which may be associated with a spin-flip or pair-hopping event) and choose the time $\tau_h$ for the annihilation operator randomly in this interval. (By next operator we always mean the neighboring operator in the direction of increasing imaginary time, taking into account periodic boundary conditions.) The distance between the inserted operators will be denoted by $l$. For the removal, we randomly pick one of the creation operators, and propose to remove the segment attached to this operator, provided it is not cut by a spin-flip or pair-hopping term. The corresponding acceptance ratio reads
\begin{align}
	&R_{\text{hyb}} (n_{\alpha, \sigma} \rightarrow n_{\alpha, \sigma} + 1) = \frac{\beta l_{\max} }{n_{\alpha, \sigma} + 1} \nonumber \\ 
	& \times e^{ (\mu - \Delta_\alpha)l -\sum_{\beta, \sigma'  \atop\ne \alpha,\sigma}  U_{\alpha, \sigma}^{\beta, \sigma'} l_{\text{overlap}}^{\beta, \sigma'} }\frac{\text{det} [ M_{\alpha, \sigma}^{(n_{\alpha, \sigma}+1)}  ]^{-1} }{\text{det} [ M_{\alpha, \sigma}^{(n_{\alpha, \sigma})} ]^{-1}},
\end{align}
where $\Delta_\alpha=\pm\Delta$ for $\alpha=1,2$ and $l_{\text{overlap}}^{\beta, \sigma'}$ denotes the total length of the overlap between the inserted segment and the segments associated with the $\beta, \sigma'$-state. 

In addition to the hybridization expansion algorithm updates, we sample the spin-flip and pair-hopping terms in $C$ using the following updates:
\begin{enumerate}
\item insertion and removal of $S ( \tau_s) S^{\dagger} (\tau'_s)$, $\tau_s>\tau'_s$,
\item insertion and removal of $S^{\dagger} ( \tau'_s ) S (\tau_s)$, $\tau'_s>\tau_s$,
\item insertion and removal of $P ( \tau_p ) P^{\dagger} (\tau'_p)$, $\tau_p>\tau'_p$,
\item insertion and removal of $P^{\dagger} ( \tau'_p ) P (\tau_p)$, $\tau'_p>\tau_p$.
\end{enumerate}
The insertion of spin-flip and pair-hopping events is only possible if the impurity is in the appropriate local state, namely the  $\ket{\downarrow, \uparrow}$, $\ket{\uparrow, \downarrow}$, $\ket{0, \uparrow\downarrow}$ and $\ket{\uparrow\downarrow, 0}$ in the order given above. 

Suppose we want to insert a spin-flip operator 
$S (\tau_s)S^{\dagger} ( \tau'_s )$ with $\tau_s>\tau'_s$. 
First, we generate 
a random imaginary time $\tau'_s$ and check whether the local state at $\tau'_s$ is the appropriate one, namely $\ket{\downarrow, \uparrow}$ in this particular case. 
If it is not the correct local state, the move is rejected. If the insertion is possible, we compute the length $\l_{\max}$ from $\tau'_s$ to the next operator and choose $\tau_s$ randomly within this interval.
In the inverse procedure, we remove an $SS^{\dagger}$ operator. To do so, we randomly select an $S^{\dagger}$ operator among the $n_{\text{sf}}$ $S^{\dagger}$ operators and remove it
together with the $S$ operator next to it (in the direction of increasing time), 
provided that there are no segments or pair-hopping operators in between. 
The same strategy is used for the insertion/removal of the $S^{\dagger} ( \tau'_s) S (\tau_s)$ operator, and for the sampling of the
pair-hopping operators.  

If we denote the length of the interval between $\tau'_s$ and $\tau_s$ by $l$, the acceptance ratios for the spin-flip and pair-hopping events become
\begin{align}
R_{SS^{\dagger}} ( n_{\text{sf}} \rightarrow n_{\text{sf}} + 1 ) &= \frac{\beta l_{\text{max}}}{n_{\text{sf}}+1} J^2, \label{eq_sf_1}\\ 
R_{S^{\dagger}S} ( n_{\text{sf}} \rightarrow n_{\text{sf}} + 1 ) &= \frac{\beta l_{\text{max}}}{n_{\text{sf}}+1} J^2, \label{eq_sf_2}\\ 
R_{PP^{\dagger}} ( n_{\text{ph}} \rightarrow n_{\text{ph}} + 1 ) &= \frac{\beta l_{\text{max}}}{n_{\text{ph}}+1} e^{ -4 \Delta l } J^2, \\
R_{P^{\dagger}P} ( n_{\text{ph}} \rightarrow n_{\text{ph}} + 1 ) &= \frac{\beta l_{\text{max}}}{n_{\text{ph}}+1} e^{ 4 \Delta l } J^2.
\end{align}
(Here, we assume that the operator on the right has the smaller time argument.)
The acceptance ratio for the two spin-flip events is the same because the intra-orbital interaction is spin symmetric, and the occupation of the orbitals does not change. 
The asymmetry in the pair-hopping case comes from the crystal field splitting, which favors the occupation of one of the orbitals.

\subsection{Retarded interactions}

As mentioned in the introduction, realistic low-energy models of correlated materials involve retarded interactions. In the case of the two-orbital model considered here, the DMFT impurity action 
thus takes the following general form
\begin{equation}\label{Eq:Seff}
S_\text{eff}=S_0+S_\text{int}+S_\text{hyb},
\end{equation}
where $S_0$ contains the chemical potential and crystal field terms, $H_\text{hyb}$ the hybridization functions and $S_\text{int}$ is given by
\begin{align}
S_\text{int}&=\frac{1}{2} \sum_{\begin{subarray}{l} \alpha, \sigma \\ \beta, \sigma' \end{subarray}} \int_0^{\beta} \dd \tau \int_0^{\beta} \dd \tau'  
	n_{\alpha, \sigma} ( \tau ) \mathcal{\tilde U}_{\alpha, \sigma}^{\beta, \sigma'} ( \tau - \tau' ) n_{\beta, \sigma'} ( \tau' )\nonumber\\
	&+\frac{1}{2} \sum_{\sigma,\sigma'} \int_0^{\beta} \dd \tau \int_0^{\beta} \dd \tau'  \mathcal{\tilde J}(\tau-\tau') \Big[X^{12}_\sigma(\tau)X^{21}_{\sigma'}(\tau')\nonumber\\
	&+X^{21}_\sigma(\tau)X^{12}_{\sigma'}(\tau')+X^{12}_\sigma(\tau)X^{12}_{\sigma'}(\tau')+X^{21}_\sigma(\tau)X^{21}_{\sigma'}(\tau')\Big],
\label{action}
\end{align}
where $X^{12}_\sigma =  c^\dagger_{1, \sigma} c_{2, \sigma}$ and $X^{21}_\sigma = c^\dagger_{2, \sigma} c_{1, \sigma}$.

The time-dependent interactions have an instantaneous component corresponding to the bare interaction and an (attractive) retarded component describing the effect of screening: $\mathcal{\tilde J}(\tau)=J_\text{bare}\delta(\tau)+J_\text{ret}(\tau)$ and $\mathcal{\tilde U}(\tau)=U_\text{bare}\delta(\tau)+U_\text{ret}(\tau)$. 
Note that the $J_\text{bare}\delta(\tau)$ contribution in the second term yields not only the instantaneous spin-flip and pair-hopping terms $\mathcal{H}_\text{sf}+\mathcal{H}_\text{sf}^\dagger+\mathcal{H}_\text{ph}+\mathcal{H}_\text{ph}^\dagger$ (with $J=J_\text{bare}$), but in addition also a same-spin inter-orbital density-density interaction $-J_\text{bare}\sum_{\alpha<\beta, \sigma} n_{\alpha,\sigma}n_{\beta,\sigma}$ and a chemical potential term $\tfrac12 J_\text{bare}\sum_{\alpha,\sigma}n_{\alpha,\sigma}$. Therefore, in the rotationally invariant case, the density-density interactions in the first term are chosen as
\begin{align}
\mathcal{\tilde U}^{\alpha,\sigma}_{\alpha,\sigma'}(\tau)&=\mathcal{\tilde U}(\tau),\\
\mathcal{\tilde U}^{1,\sigma}_{2,\sigma'}(\tau)=\mathcal{\tilde U}^{2,\sigma}_{1,\sigma'}(\tau)&=\mathcal{\tilde U}(\tau)-2\mathcal{\tilde J}(\tau). 
\end{align}

The retarded interactions $U_{\rm ret}(\tau), J_{\rm ret}(\tau)$ arise from some electron-boson coupling term of the form 
\begin{eqnarray}
 {\mathcal H}_\text{e-b}  = \sum_{\nu}\sum_{\alpha,\alpha',\sigma} \lambda^{\nu}_{\alpha,\alpha'}  c^\dagger_{\alpha, \sigma} c_{\alpha', \sigma} (b_\nu + b^{\dagger}_{\nu}), 
\end{eqnarray}
where $b^{\dagger}_{\nu}$ ($b_{\nu}$) denotes the creation (annihilation) operator for the $\nu$th bosonic degree of freedom.
The boson one-body part is given by $\sum_{\nu} \omega_{\nu} b^{\dagger}_{\nu} b_{\nu}$.
Here, the bosonic degrees of freedom represent plasmons, bosonic modes corresponding to single-particle excitations, 
and phonons. 
The former two are responsible for the dynamical screening processes which are taken into account in the down-folding procedure. 
The phonons further reduce the resulting interactions values. 

In terms of $\lambda^{\nu}_{\alpha,\alpha'}$ and $\omega_{\nu}$, $U_{\rm ret}(\omega)$ are $J_{\rm ret}(\omega)$ can be written as 
\begin{align}
U_{\rm ret } (\omega) &= \sum_{\nu } \frac{2  (\lambda^{\nu}_{1,1} )^2} { \omega^2 - \omega_{\nu}^2} = \sum_{\nu } \frac{2  (\lambda^{\nu}_{2,2} )^2} { \omega^2 - \omega_{\nu}^2}, \label{Eq:Uret} \\ 
J_{\rm ret } (\omega) &= \sum_{\nu } \frac{2  ( \lambda^{\nu}_{1,2} )^2 } { \omega^2 - \omega_{\nu}^2} = \sum_{\nu } \frac{2  ( \lambda^{\nu}_{2,1} )^2 } { \omega^2 - \omega_{\nu}^2} 
 =  \sum_{\nu } \frac{2  \lambda^{\nu}_{1,2}  \lambda^{\nu}_{2,1}  } { \omega^2 - \omega_{\nu}^2}. \label{Eq:Jret}
\end{align} 
Nowadays, {\it ab initio} estimates of these retarded interactions can be obtained by the constrained random phase approximation (cRPA)\cite{Aryasetiawan_2004} for 
the electronic screening contribution, and by the constrained density-functional perturbation theory (cDFPT)~\cite{PhysRevLett.112.027002,note_phonon_downfold}  for the phonon contribution.
In reality, there is a continuum of electronic screening frequencies, so that the sum in Eqs.~(\ref{Eq:Uret}) and (\ref{Eq:Jret}) becomes an integral over frequencies.\cite{Werner_2012} 

In most strongly correlated materials, the screened $U$ and $J$ remain positive, and the frequency dependence of $J$ is rather weak.\cite{Sasioglu2011}  
An interesting exception are the alkali-doped fullerides, where the Wannier functions are delocalized on a C$_{60}$ molecule. Here, the bare exchange interaction $J_\text{bare}$ becomes small ($J_\text{bare} \sim 0.1$ eV, bandwidth $\sim 0.5$ eV).\cite{PhysRevB.85.155452} The electronic screening contributions reduce the static value of $J$ to $\sim 0.035$ eV, which is still positive.  The important low-energy screening contributions come from phonons, which yield an attraction of $\sim -0.05$ eV, inverting the sign of the static exchange interaction.\cite{nomura_C60_paper}
The resulting multiorbital Hamiltonian with an overscreened $J$ is of great interest, since a negative $J_\text{scr}$ can induce nontrivial synergies between the correlated electrons and phonons, leading to an exotic $s$-wave superconductivity.\cite{Capone_2002,RevModPhys.81.943,nomura_C60_paper} 
However, the effect of the dynamical screening of $J$ is still an open issue. 
The algorithm presented in this paper provides a basis for attacking this challenging problem. 

In solving the impurity model with the action $S_{\rm eff}$ in Eq.~(\ref{Eq:Seff}), we treat the density-density and spin-flip/pair-hopping terms in Eq.~(\ref{action}) separately. It is therefore convenient to shift the $J_\text{bare}\delta(\tau)\delta_{\sigma,\sigma'}$ contribution to the density-density term and 
to write the action $S_\text{int}$ with the interactions $\tilde {\mathcal J}$ and $\tilde {\mathcal U}$ replaced by  
$\mathcal{J}(\tau)=J_\text{bare}\delta(\tau)\delta_{\sigma,-\sigma'}+J_\text{ret}(\tau)$ and $\mathcal{U}(\tau)=U_\text{bare}\delta(\tau)-J_\text{bare}\delta(\tau)\delta_{\sigma,\sigma'}(1-\delta_{\alpha,\beta})+U_\text{ret}(\tau)$. In this formulation, the density-density interactions become
\begin{align}
\mathcal{U}^{\alpha,\sigma}_{\alpha,\sigma'}(\tau)&=U_\text{bare}\delta(\tau)+U_\text{ret}(\tau),\\
\mathcal{U}^{1,\sigma}_{2,-\sigma}(\tau)=\mathcal{U}^{2,\sigma}_{1,-\sigma}(\tau)&=(U_\text{bare}-2J_\text{bare})\delta(\tau)\nonumber\\
&\hspace{10mm}+(U_\text{ret}(\tau)-2J_\text{ret}(\tau)), \\
\mathcal{U}^{1,\sigma}_{2,\sigma}(\tau)=\mathcal{U}^{2,\sigma}_{1,\sigma}(\tau)&=(U_\text{bare}-3J_\text{bare})\delta(\tau)\nonumber\\
&\hspace{10mm}+(U_\text{ret}(\tau)-2J_\text{ret}(\tau)).
\end{align}

The half-filling condition, which is $\mu_{1/2}=\tfrac{3}{2}U_\text{bare}-\tfrac{5}{2}J_\text{bare}$ in the model without dynamical screening, becomes $\mu_{1/2}=\tfrac{3}{2}U_\text{scr}-2J_\text{scr}-\tfrac12 J_\text{bare}$, with $U_\text{scr}=U(\omega=0)$ and $J_\text{scr}=J(\omega=0)$. 
In the limit $J_\text{ret}(\tau)=(J_\text{scr}-J_\text{bare})\delta(\tau)$ (high frequency screening) we recover the rotationally invariant $S_\text{int}$ with static interactions equal to the screened values. (Again, one has to take into account the additional density-density and chemical potential terms resulting from the $X^{12}X^{21}$ and $X^{21}X^{12}$ operators.) 

\begin{figure}
\includegraphics[width=9cm]{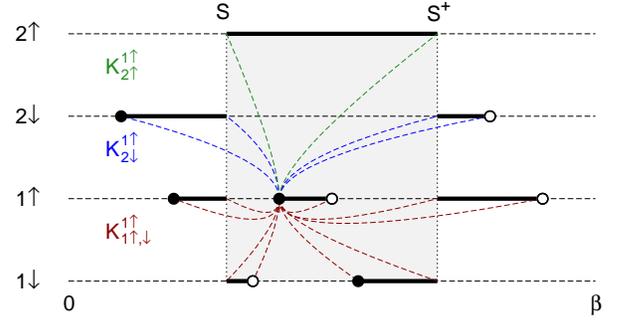}
\caption{\label{fig:retardedu} 
(Color online) 
Illustration of an order $n_{\textrm{hyb}} = 4$, $n_{\textrm{sf}} = 1$ configuration for a model with retarded density-density interactions. 
Dashed lines indicate interactions $K^{\beta,  \sigma'}_{\alpha, \sigma} ( \tau )$ connecting all
pairs of hybdrization, spin-flip and pair-hopping events (only the interactions involving the creation operator of the non-cut segment are shown). The sign $s_is_j$ associated with each dashed line is $\pm 1$ depending on the types of operators (creation/annihilation) which it connects.  
}
\end{figure}

\subsubsection{Retarded density-density interactions}
\label{sec_ret_dens}

The density-density contribution to $S_\text{int}$ can be calculated efficiently using the technique discussed in Refs.~\onlinecite{Werner_2007, Werner_2010, Ayral_2013}. The retarded part leads to an additional  
weight factor $w_{\text{screen}} ( \{ \tau_i \} )$ of the form 
\begin{equation}
	w_{\text{screen}} ( \{ \tau_i \} ) = \exp \left( \sum_{2n \geq i > j \geq 1} s_i^{\alpha,  \sigma} s_j^{\beta,  \sigma'} K^{\alpha,  \sigma}_{\beta, \sigma'} ( \tau_i - \tau_j ) \right),\label{wscreen}
\end{equation}
where $n = n_h+n_\mathrm{sf}+n_\mathrm{ph}$, the $\tau_i$ are the times corresponding to segment start- or end-points (which can be the locations of hybridization, spin-flip or pair-hopping operators), and
$s=\pm 1$ is a sign ($+1$ for segment start-points and $-1$ for segment end-points).  The function $K^{\alpha,\sigma}_{\beta,\sigma'}(\tau)$  is obtained from the twice integrated retarded interaction $U_\text{ret}(\tau)$ and $J_\text{ret}(\tau)$,\cite{Ayral_2013} and thus is also orbital- and spin-dependent: 
\begin{align}
	K^{\alpha,  \sigma}_{\alpha, \sigma'} ( \tau ) &= K_U ( \tau ),\\
	K^{1,  \sigma}_{2, \sigma'} ( \tau ) = K^{2,  \sigma}_{1, \sigma'} ( \tau ) &= K_U ( \tau ) - 2 K_J ( \tau ). 
\end{align}
with $K''_U(\tau)=U_\text{ret}(\tau)$ and $K''_J(\tau)=J_\text{ret}(\tau)$. 
Both $K_U(\tau)$ and $K_J(\tau)$ are defined in the range $0 < \tau < \beta$, are $\beta$-periodic and symmetric around $\tau = \beta/2$, and satisfy $K_{U,J}(0^+) = K_{U,J} (\beta^-) = 0$. 

The structure of the configurations thus remains the same as in the simulations without retarded density-density interactions (they consist of a collection of hybridization, spin-flip and pair-hopping events), the only difference is that now each pair of creation/annihilation operators (both hybridization events and segment start- or end-points corresponding to spin flip and pair hopping events) are linked by lines representing the ``interaction" $s_i^{\alpha,  \sigma} s_j^{\beta,  \sigma'} K^{\alpha,  \sigma}_{\beta, \sigma'} ( \tau_i - \tau_j )$, see Fig.~\ref{fig:retardedu}. In a local update, only the lines connected to the inserted or removed operators have to be considered. 

More explicitly, the acceptance ratio for a spin-flip (or pair hopping) insertion 
becomes 
\begin{align}
	R_{SS^{\dagger}} &\propto  e^{ \sum_i'\sum_{j} s_i^{\alpha,  \sigma} s_j^{\beta,  \sigma'} K^{\alpha,  \sigma}_{\beta, \sigma'} ( \tau_i - \tau_j ) }, \label{eq:sfaccrat} \\ 
	R_{PP^{\dagger}} &\propto  e^{ \sum_i'\sum_{j} s_i^{\alpha,  \sigma} s_j^{\beta,  \sigma'} K^{\alpha,  \sigma}_{\beta, \sigma'} ( \tau_i - \tau_j ) }, \label{eq:phaccrat}
\end{align}
where the sum over $i$ runs over all operators associated with the new spin-flip (or pair-hopping) event, and the sum over $j$ runs over all the other operators in the configuration ($i$,$j$ pairs in the same orbital can be ignored).

\subsubsection{Retarded spin-flip and pair-hopping terms}

\begin{figure}[t]
\includegraphics[width=9cm]{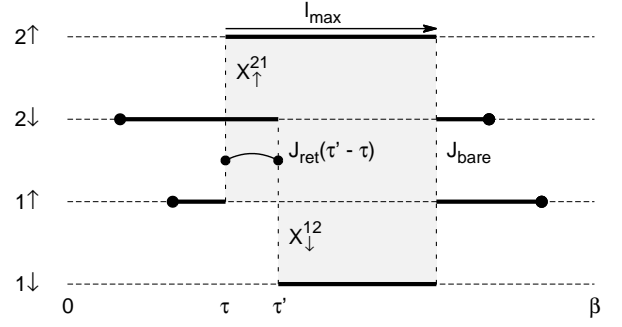}
\caption{\label{fig:retardedj1} 
Retarded spin-flip insertion. An instantaneous spin-flip operator $S(\tau)=X^{12}_\downarrow(\tau)X^{21}_\uparrow(\tau)$ is split into two separate hopping events $X^{21}_\uparrow(\tau)$ and $X^{12}_\downarrow(\tau')$ by randomly choosing the position of one of the operators (here $X^{12}_\downarrow$) in the interval of length $l_\text{max}$.
}
\end{figure}

The sampling of the retarded spin-flip or pair-hopping operators is analogous to the algorithm discussed in Ref.~\onlinecite{Otsuki_2013}, i.e. we expand the partition function in powers of these terms and sample their contribution stochastically.  
In order to insert a retarded spin-flip operator, we replace an instantaneous spin-flip event by a retarded $X^{12}_{\sigma}(\tau) X^{21}_{\sigma'}(\tau')$  
operator, with $\tau\ne\tau'$ and $\sigma\ne\sigma'$.
We describe the procedure here for retarded spin flips (the retarded pair hoppings are sampled in the same manner).    
First, we randomly select the type of operator ($S$ or $S^\dagger$) which we want to split. In the following, we assume it is an $S$ operator. Next, we choose one of the $n_{\text{sf}}^S$ instantaneous spin-flip events corresponding to $S$ operators. Suppose we select the $S$ operator located at time $\tau$.
This operator can be written as $S(\tau)=X^{12}_\downarrow(\tau)X^{21}_\uparrow(\tau)$. We randomly select either the $X^{12}_\downarrow$ or $X^{21}_\uparrow$ operators for the proposed
shift on the time axis (suppose it is $X^{12}_\downarrow$). To fix the new location, we compute the distance $l_\text{max}$ to the next operator in the forward-direction (taking periodic boundary conditions into account) and choose the time $\tau'$ randomly within the interval of length $l_\text{max}$. The proposed move is from the instantaneous spin-flip event with weight $-J_\text{bare}X^{12}_\downarrow(\tau)X^{21}_\uparrow(\tau)d\tau$ to the retarded spin flip with weight $-J_\text{ret}(\tau'-\tau)X^{12}_\downarrow(\tau')X^{21}_\uparrow(\tau)d\tau^2$. In this new configuration, the hopping operators are connected by the interaction line $J_\text{ret}(\tau'-\tau)=\mathcal{J}(\tau'-\tau)$ (see Fig.~\ref{fig:retardedj1}).

In the inverse move, we remove the retarded spin-flip event by randomly selecting one of the $n_{\text{rsf}}^{S}$ retarded spin-flip pairs corresponding to $S$. 
We then randomly choose one of the operators and try to shift the other operator to its position on the time-axis. If there are other operators between the retarded spin-flip operators, the move is rejected. With these procedures, the acceptance ratio for the retarded spin-flip insertion becomes
\begin{align}
	&R_{XX} (n_\text{sf}^S, n_{\text{rsf}}^S \rightarrow n_\text{sf}^S-1,n_{\text{rsf}}^S + 1) = \nonumber\\
	&\hspace{20mm}\frac{n_{\text{sf}}^S  l_{\max} }{n_{\text{rsf}}^S + 1} 
	e^{ -\sum  U_{\alpha, \sigma}^{\beta, \sigma'} l_{\text{overlap}}^{\beta, \sigma'} }\frac{ J_\text{ret}(\tau'-\tau)}{J_\text{bare}}.\label{RXX}
\end{align}
Because the number of instantaneous spin-flip events $n_\text{sf}^S$ associated with $S$ operators can now be different from the number $n_\text{sf}^{S^\dagger}$ associated with $S^\dagger$ operators, we have to keep track of these perturbation orders separately. In the instantaneous spin-flip updates, one then uses $n_\text{sf}^{S^\dagger}$ in Eq.~(\ref{eq_sf_1}) and $n_\text{sf}^S$ in Eq.~(\ref{eq_sf_2}). 
We also note that in the usual situation where $J_\text{bare}>0$ and $\mathcal{J}(\tau)<0$ ($0<\tau<\beta$), Eq.~(\ref{RXX}) implies that the Monte Carlo sampling for the simulation with retarded spin-flip and pair-hopping terms will suffer from a sign problem. 

If the screening frequency is high, so that $J_\text{ret}(\tau)$ approaches a $\delta$-function, it is more efficient to absorb this factor into the proposal probability. More specifically, we propose the time $\tau'$ in the interval of length $l_\text{max}$ according to the probability distribution $-J_\text{ret}(\tau'-\tau)/\int_\tau^{\tau+l_\text{max}}d\tau'|J_\text{ret}(\tau'-\tau)|$. In this case, the ratio of acceptance probabilities becomes
 \begin{align}
	&R_{XX} (n_\text{sf}^S, n_{\text{rsf}}^S \rightarrow n_\text{sf}^S-1,n_{\text{rsf}}^S + 1) = \nonumber\\
	&\hspace{0mm}\frac{n_{\text{sf}}^S }{n_{\text{rsf}}^S + 1} 
	e^{ -\sum  U_{\alpha, \sigma}^{\beta, \sigma'} l_{\text{overlap}}^{\beta, \sigma'} }\frac{ \int_\tau^{\tau+l_\text{max}}d\tau'|J_\text{ret}(\tau'-\tau)|}{J_\text{bare}}(-1).\label{RXX_efficient}
\end{align}

\begin{figure}[t]
\includegraphics[width=9cm]{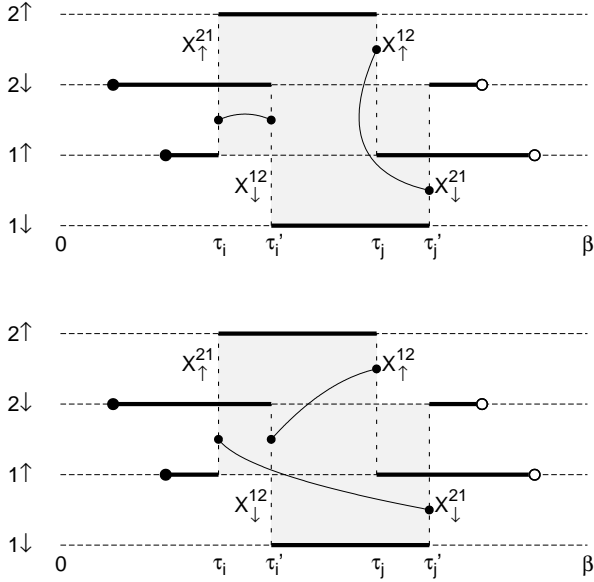}
\caption{\label{fig:retardedj2} 
Swap updates. We can generate additional configurations with retarded pairs of $X$ operators by swapping the end-points of two retarded spin-flip or pair-hopping events.
}
\end{figure}

\subsubsection{Kink updates}

Because of the retarded $X_\sigma X_\sigma$ terms in Eq.~(\ref{action}), an ergodic sampling requires additional updates. 
One of these additional updates swaps the $J_\text{ret}$-links of two randomly chosen pairs of retarded $X$-operators (Fig. \ref{fig:retardedj2}). If the time points corresponding to these retarded pairs are $(\tau_i, \tau_i')$ and $(\tau_j, \tau_j')$, and the configuration before and after the swapping is denoted by $C$ and $C'$, respectively, the acceptance probability for the move is given by 
\begin{equation}
R_\text{swap}( C \rightarrow C' ) = \frac{ J_\text{ret}( \tau_i' - \tau_j ) J_\text{ret}( \tau_j' - \tau_i ) }{ J_\text{ret}( \tau_i' - \tau_i) J_\text{ret}( \tau_j' - \tau_j ) }.
\end{equation}
In particular, this type of update allows us to produce configurations with retarded interactions of the type $-J_\text{ret}(\tau'-\tau)X^{12}_\sigma(\tau') X^{12}_\sigma(\tau)$ and $-J_\text{ret}(\tau'-\tau)X^{21}_\sigma(\tau') X^{21}_\sigma(\tau)$.

We furthermore need the insertion/removal of ``kinks" $-J_\text{ret}(\tau'-\tau)X^{12}_\sigma(\tau') X^{21}_\sigma(\tau)$. Suppose we have $n_\text{kink}^K$ kinks of type $K$ and we randomly choose one type for the insertion or removal. For the insertion, we randomly select the time $\tau$ for the first operator. If the kink insertion is possible, we compute the length $l_\text{max}$ of the interval in which the second operator can be inserted and place it at the time $\tau'$ according to the distribution $-J_\text{ret}(\tau'-\tau)/\int_\tau^{\tau+l_\text{max}}d\tau'|J_\text{ret}(\tau'-\tau)|$. In the reverse move, we randomly select one of the $n_\text{kink}^K+1$ kinks of type $K$, and remove the corresponding $X$ operators if there is no other operator in between. The corresponding ratio of acceptance probabilities is 
\begin{align}
	&R_{K} (n_{\text{kink}}^K \rightarrow n_\text{kink}^K+1) = \nonumber\\ 
	&\hspace{0mm}\frac{\beta}{n_{\text{kink}}^K + 1} 
	e^{ -\sum  U_{\alpha, \sigma}^{\beta, \sigma'} l_{\text{overlap}}^{\beta, \sigma'} } \int_\tau^{\tau+l_\text{max}}d\tau'|J_\text{ret}(\tau'-\tau)|.\label{RK}
\end{align}

\subsection{Simplified model}

To avoid the sign problem originating from configurations with an odd number of retarded spin-flips/pair-hoppings, it is useful to consider a simplified action which has a retarded density-density interaction, but only instantaneous spin-flip and pair-hopping terms: 
\begin{align}
S_\text{int}^\text{simp}&=\frac{1}{2} \sum_{\begin{subarray}{l} \alpha, \sigma \\ \beta, \sigma' \end{subarray}} \int_0^{\beta} \!\!\dd \tau \!\int_0^{\beta} \!\!\dd \tau'  
	n_{\alpha, \sigma} ( \tau ) (\mathcal{U}_s)_{\alpha, \sigma}^{\beta, \sigma'} ( \tau - \tau' ) n_{\beta, \sigma'} ( \tau' )\nonumber\\
	&+ \int_0^{\beta} \dd\tau J_\text{scr} \Big[S(\tau)+S^\dagger(\tau)+P(\tau)+P^\dagger(\tau)\Big].
\label{action_simplified}
\end{align}
Note that we choose here the screened Hund coupling $J_\text{scr}=\mathcal{J}(\omega=0)$ in front of the second term to correctly capture the limit of high screening frequency. The retarded density-density interaction of the simplified model is
\begin{align}
(\mathcal{U}_s)^{\alpha,\sigma}_{\alpha,\sigma'}(\tau)&=U_\text{bare}\delta(\tau)+U_\text{ret}(\tau),\\
(\mathcal{U}_s)^{1,\sigma}_{2,-\sigma}(\tau)=(\mathcal{U}_s)^{2,\sigma}_{1,-\sigma}(\tau)&=(U_\text{bare}-2J_\text{bare})\delta(\tau)\nonumber\\
&\hspace{0mm}+(U_\text{ret}(\tau)-2J_\text{ret}(\tau)), \\
(\mathcal{U}_s)^{1,\sigma}_{2,\sigma}(\tau)=(\mathcal{U}_s)^{2,\sigma}_{1,\sigma}(\tau)&=(U_\text{bare}-3J_\text{bare})\delta(\tau)\nonumber\\
&\hspace{0mm}+(U_\text{ret}(\tau)-3J_\text{ret}(\tau)),
\end{align} 
and the $K$-functions in Eq.~(\ref{wscreen}) become 
\begin{align}
	K^{\alpha,\sigma}_{\alpha, \sigma'} ( \tau ) &= K_U ( \tau ),\\
	K^{1,  \sigma}_{2, -\sigma} ( \tau ) = K^{2,  \sigma}_{1, -\sigma} ( \tau ) &= K_U ( \tau ) - 2 K_J ( \tau ),\\
	K^{1,  \sigma}_{2, \sigma} ( \tau ) = K^{2, \sigma}_{1, \sigma} ( \tau ) &= K_U ( \tau ) - 3 K_J ( \tau ).
\end{align}
The half-filling condition for the simplified model is $\mu_{1/2}=\tfrac32 U_\text{scr}-\tfrac52 J_\text{scr}$. 

Because of the spin-dependence of the inter-orbital interaction, Eqs.~(\ref{eq:sfaccrat}) and (\ref{eq:phaccrat}) now read
\begin{align}
	R_{SS^{\dagger}} &\propto  e^{ -4 K_J (l) + \sum_i'\sum_{j} s_i^{\alpha,  \sigma} s_j^{\beta,  \sigma'} K^{\alpha,  \sigma}_{\beta, \sigma'} ( \tau_i - \tau_j ) }, \\ 
	R_{PP^{\dagger}} &\propto  e^{ -20 K_J (l) + \sum_i'\sum_{j} s_i^{\alpha,  \sigma} s_j^{\beta,  \sigma'} K^{\alpha,  \sigma}_{\beta, \sigma'} ( \tau_i - \tau_j ) }, 
\end{align}
where $l$ is the distance between the inserted spin-flip or pair-hopping operators. Apart from this change, the simulation proceeds as discussed previously for the case of retarded density-density interactions. 

\begin{figure}[t]
	\includegraphics[width=8.5cm]{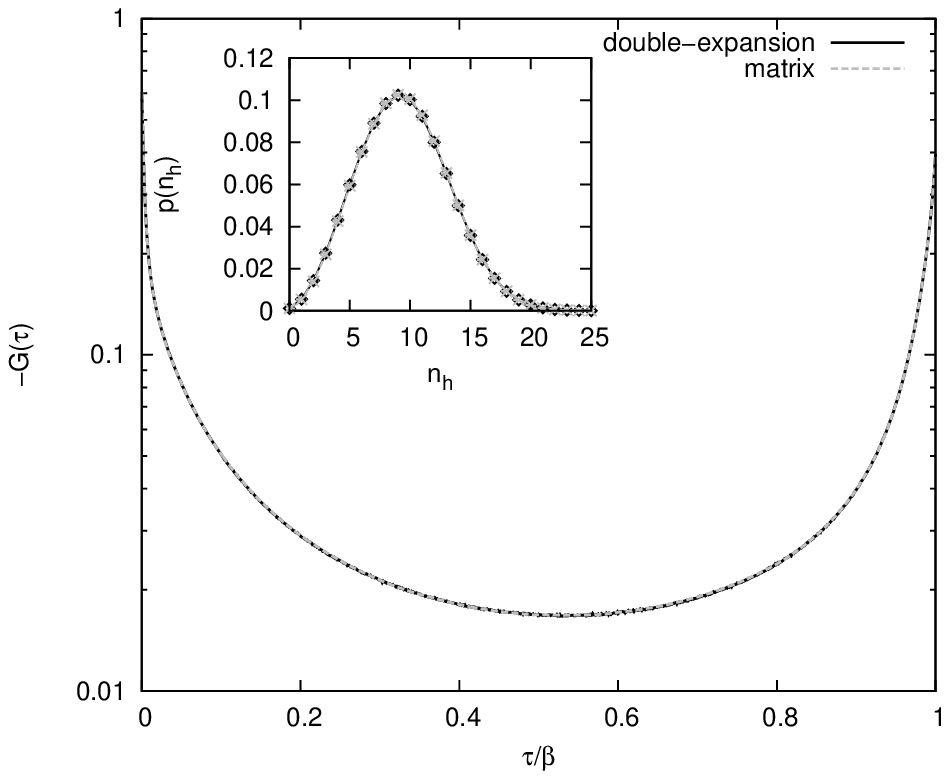}
	\includegraphics[width=8.5cm]{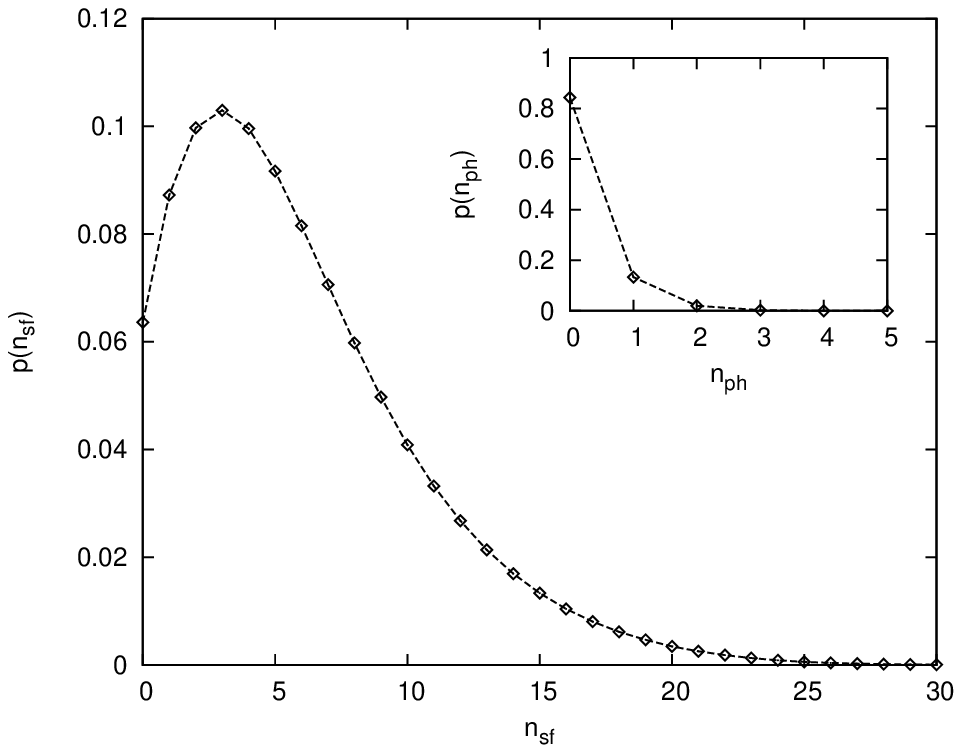}
	\caption{\label{fig_gfcomp} 
	Comparison of simulations results for $U = 8$, $J = 8/6$, $\mu = 4.5$ and $\beta = 50$ (filling $n$=0.396). The top panel shows the Green's function, with the solid line showing the result from the double-expansion solver, while the dashed line has been obtained with the matrix formalism. The inset in the top panel shows the distribution of hybridization orders, which is identical in both methods. The bottom panel shows the distribution of spin-flip and, as an inset, the distribution of pair hopping orders for this parameter set. 
}
\end{figure}

\section{Tests of the solver}\label{sec:tests}

We first show the results of some tests of the double expansion solver, starting with a model that contains only instantaneous spin-flip and pair-hopping terms, and a static $U$. The top panel of Fig.~\ref{fig_gfcomp} compares the Green's functions obtained with the new solver and with the matrix formalism\cite{Werner_2006_Matrix} for a half-filled orbitally-degenerate model with $U=8$, $J=1.33$, $\mu=4.5$ and $\beta=50$ (DMFT solution for a semi-circular density of states with bandwidth $4$). Both results agree within statistical errors (the error bars are comparable to the line thickness), and as shown in the inset, also the distribution of the perturbation orders $n_h$ is identical. This is because the sampling of the hybridization operators is independent of the treatment of the spin-flip and pair-hopping terms (exact treatment in the time-evolution $e^{-\tau H_\text{loc}}$ in the case of the matrix formalism versus stochastic sampling in the double-expansion approach). Obviously, also the physical quantities derived from this distribution of perturbation orders, such as the kinetic energy,\cite{Haule_2007} will agree. 

\begin{figure}
	\includegraphics{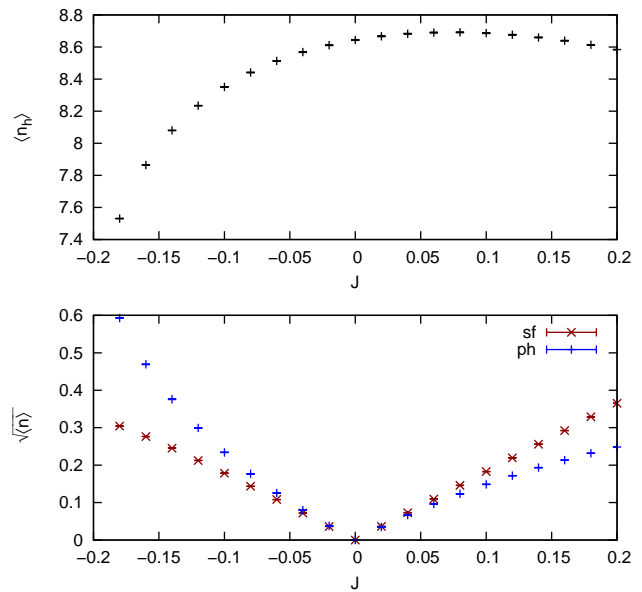}
	\caption{\label{fig_njscan} 
	Average perturbation order for $n_h$, $n_\text{sf}$, and $n_\text{ph}$ as a function of $J$ for $U=3$ and half-filling ($\beta=25$).  
}
\end{figure}

The bottom panel shows the distribution of spin-flip orders $n_\text{sf}$ and as an inset the distribution of pair-hopping orders $n_\text{ph}$. In this simulation, configurations with spin-flip orders up to about $25$ are relevant, while configurations with more than 3 pair-hopping events are rarely generated. The average spin-flip order is $5.93$ whereas the average pair-hopping order is only $0.18$. This is because a positive $J$ favors high-spin states ($S=1$), and one of the high-spin states, $\frac{1}{\sqrt{2}} ( \ket{ \uparrow \downarrow } + \ket{\downarrow \uparrow} )$, 
is an eigenstate of  $(S+S^\dagger)$. For negative $J$, the situation is opposite and the average perturbation order for the pair-hopping term becomes large. 

The scaling of the perturbation orders with $J$ is illustrated in Fig.~\ref{fig_njscan} for a half-filled metallic system. For small $|J|$, $n_\text{sf}$ and $n_\text{ph}$ grows roughly proportional to $J^2$. This result is expected since the $J$-terms are treated by a weak-coupling expansion, and in the simulation with static $J$, each $S$ or $P$ operator must be balanced by a $S^\dagger$ or $P^\dagger$. Hence, the weight (\ref{eq:weight}) is proportional to $(J^2)^{n_\text{sf}}$ and $(J^2)^{n_\text{ph}}$. At larger $|J|$, deviations from the quadratic behavior appear, due to changes in the hybridization order and interference between spin-flip and pair-hopping terms. We also notice that $n_\text{sf}$ grows more rapidly than $n_\text{ph}$ on the $J>0$ side, while it is the opposite on the $J<0$ side. As mentioned above, this is because of the increased weight of high-spin (low-spin) states in the model with $J>0$ ($J<0$). 

We next test the implementation with retarded interactions. As was already mentioned, in the case of a retarded $\mathcal{J}(\tau)$, the expansion in the retarded spin-flip and pair-hopping terms produces a sign problem. Therefore, we first consider the simplified model (\ref{action_simplified}) where only the density-density interaction is retarded, whereas the spin-flip and pair-hopping terms are instantaneous. For this test, we assume a static interaction $U$, so that the retarded density-density part arises from the retardation of $J(\tau)$. The same-spin density-density interaction thus becomes $U-3\mathcal{J}(\tau)$, implying $K^{\alpha,\sigma}_{\beta,\sigma}(\tau)=-3K_J(\tau)$ for $\alpha\ne\beta$, and the half-filling condition is $\mu_{1/2}=\tfrac32 U-\tfrac52 J_\text{scr}$. 

For the frequency-dependence, we take a single-boson model with a screening frequency $\omega_J$ and a coupling strength $\lambda_J$, which corresponds to 
\begin{align}
\mathcal{J}(\omega)&=J_\text{bare}+\frac{2\lambda_J^2\omega_J}{\omega^2-\omega_J^2},
\label{singleboson}
\end{align}
and hence $J_\text{scr} = J_\text{bare}-\frac{2\lambda_J^2}{\omega_J}$. In Fig.~\ref{fig_gfcomp} we show results for $U=4$, $\mu=\mu_{1/2}$, $J_\text{bare}=1$, fixed $J_\text{scr}=0.4$ and different values of $\omega_J$. In the limit of large $\omega_J$, the results should approach those for the frequency independent parameters $U$ and $J_\text{scr}$ (see inset of the bottom panel). Indeed, as seen in the top panel of Fig.~\ref{fig_gfcomp}, the Green's function approaches the result for a static $J_\text{scr}$ as $\omega_J\rightarrow \infty$, and the same is true for the imaginary part of the self-energy at the lowest Matsubara frequency $\omega_1=\pi/\beta$ (inset). Two-particle quantities, such as the equal-time spin-correlation function $\langle S_z^2\rangle$ (middle panel) or the double occupancy $\langle n_{i\uparrow}n_{i\downarrow}\rangle$ (bottom panel) also approach the correct values in the limit of a large screening frequency. 

\begin{figure}
	\includegraphics[width=8.3cm]{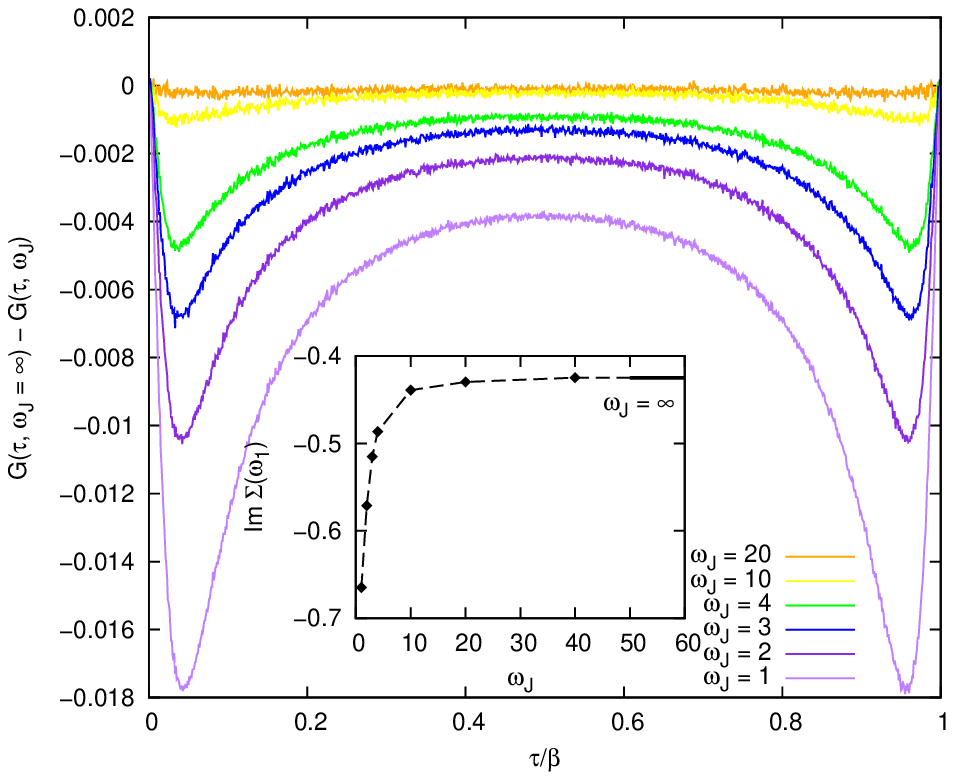}
	\includegraphics[width=8.3cm]{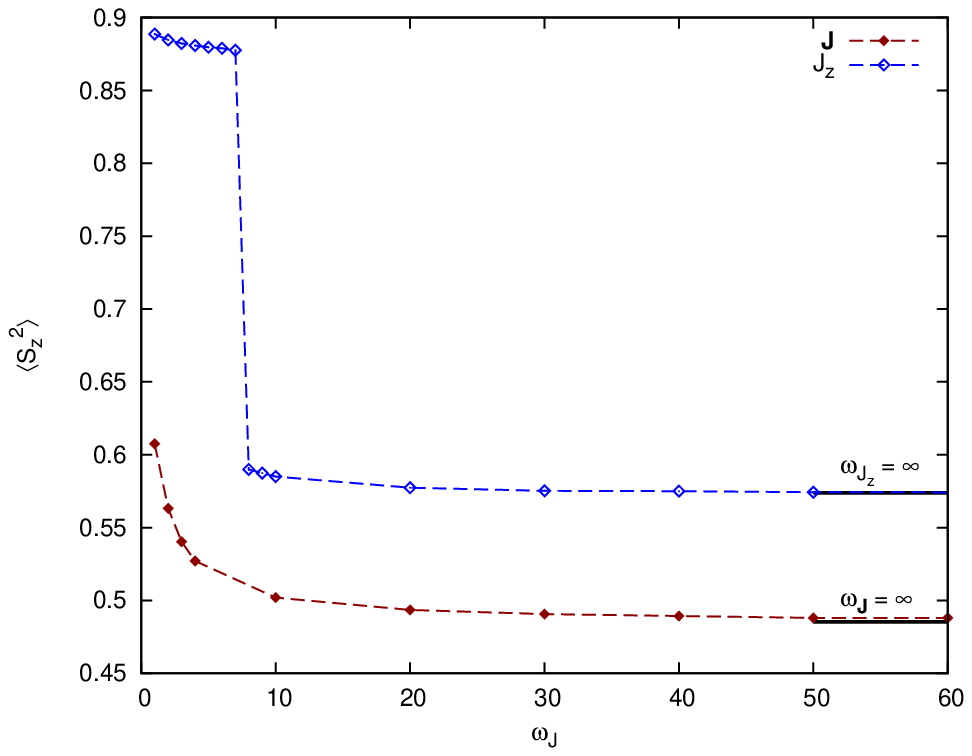}
	\includegraphics[width=8.3cm]{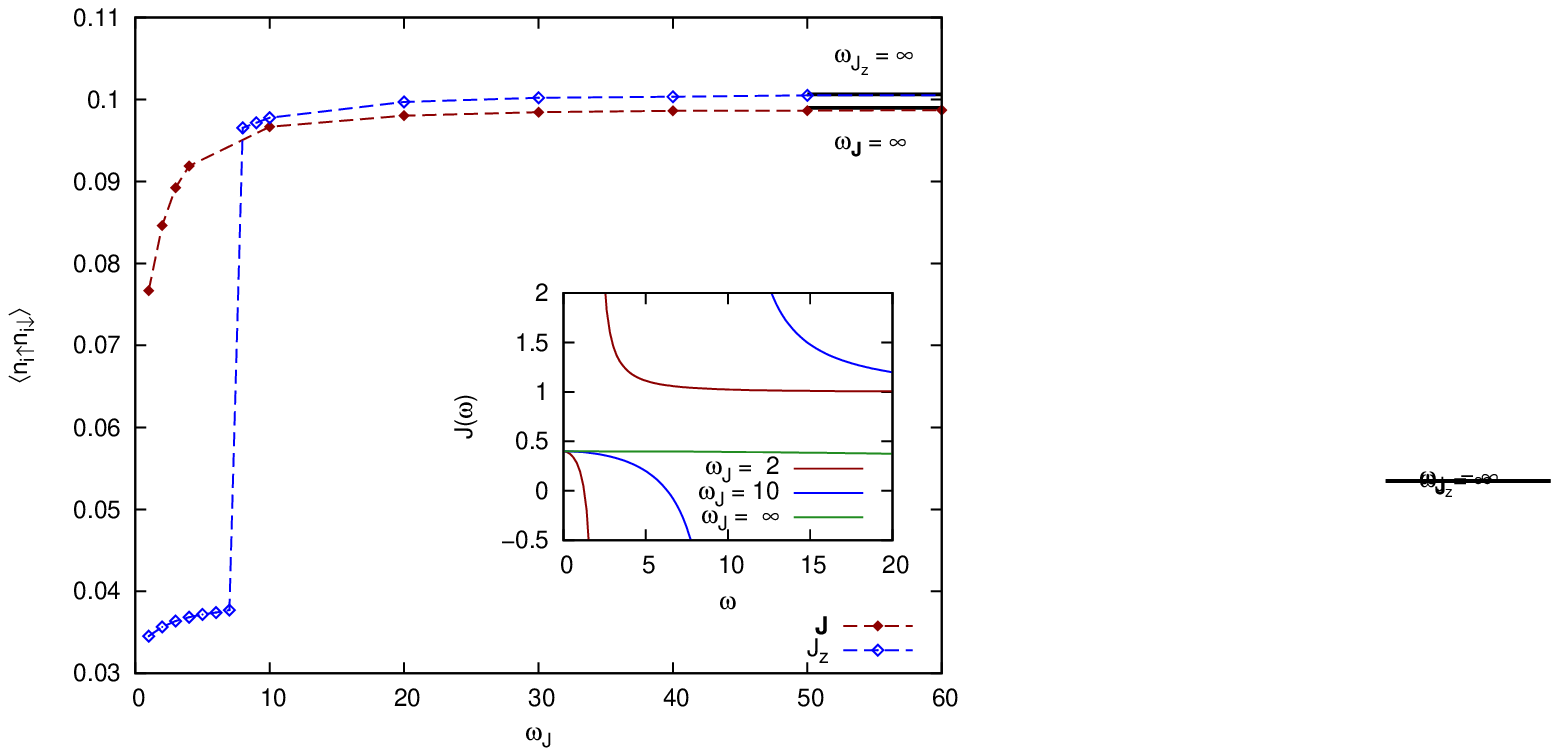}
	\caption{\label{fig_gfcomp} 
	Green's functions and local observables for $U = 4$, $J_\text{bare}=1$, $J_{\text{scr}} = 0.4$, and different screening frequencies $\omega_J$ ($\mu = \mu_{1/2}$ and $\beta = 25$). Results for the static limit $\omega_J\rightarrow \infty$ ($U=4$, $J=J_\text{scr}=0.4$) are also shown (horizontal bars in the bottom two panels) or subtracted (top panel). The curves labeled by $J$ are from simulations which include instantaneous spin-flip and pair-hopping terms (with $J=J_\text{scr}$), while the curves labeled by $J_z$ are for the density-density approximation. Insets: Im$\Sigma$ at the lowest Matsubara frequency (top panel), and $J(\omega)$ on the real-frequency axis (bottom panel).
}
\end{figure}

Besides the calculations with instantanteous spin-flip and pair-hopping terms, we also show the result from the simple density-density approximation, which ignores the spin-flip and pair-hopping contributions and treats only the (dynamical) $J_z$.  As is evident from the plots, these two approximations can produce quite different results. For example, the density-density calculation exhibits a transition to an insulating high-spin state at screening frequency $\omega_J\approx 7$, whereas the calculation with instantanteous spin-flip and pair-hopping terms yields a metallic solution down to $\omega_J=1$.

\begin{table}[b]
\caption{\label{tab:signscr}
Average signs and perturbation orders for a simulation with instantaneous spin-flips and retarded $XX$ operator pairs, $U=4$, $J_\text{bare}=1$ and $J_\text{scr}=0$. $n_\text{inst}$ is the average number of spin-flip ($S$ or $S^\dagger$) operators, while $n_\text{ret}$ is the average number of $XX$ operators linked by $J_\text{ret}(\tau)$. $n_\text{kink}$ corresponds to the subset of kink-operators $X^{\alpha\beta}_\sigma X^{\beta\alpha}_\sigma$.
}
\begin{ruledtabular}
\begin{tabular}{cc cc cc}
$\omega_J$ & $\beta$ & sign & $n_\text{inst}$ & $n_\text{ret}$ & $n_\text{kink}$\\ 
\hline
2  &  1 & 0.6768 & 0.379 & 1.106 & 0.451    \\ % 
2  &  2 & 0.2807 & 1.139 & 2.805 & 0.821    \\ % 
2  &  3 & 0.1036 & 1.913 & 4.472 & 1.108    \\ % 
2  &  4 & 0.0366 & 2.665 & 6.015 & 1.326    \\ % 
2  &  5 & 0.0127 & 3.397 & 7.455 & 1.500    \\ % 
\hline                               
10 &  1 & 0.6369 & 0.432 & 1.263 & 0.501    \\ % 
10 &  2 & 0.2093 & 1.388 & 3.252 & 0.918    \\ % 
10 &  3 & 0.0623 & 2.363 & 5.062 & 1.206    \\ % 
10 &  4 & 0.0196 & 3.290 & 6.760 & 1.421    \\ % 
10 &  5 & 0.0064 & 4.175 & 8.426 & 1.601    \\ % 
\end{tabular}
\end{ruledtabular}
\end{table}

Finally, let us quantify the effect of the retarded spin-flips and kinks. To simplify the calculations, we switch off the pair-hopping terms and show the results after one iteration starting from a metallic solution (noninteracting hybridization function for $\beta=50$). We consider a model with $U=4$, $J_\text{bare}=1$ and $J_\text{scr}=0$. 
 Table~\ref{tab:signscr} shows the average order of instantaneous and retarded $XX$ pairs, $n_\text{inst}$ and $n_\text{ret}$, as well as the average sign, for screening frequencies $\omega_J=2$ and $\omega_J=10$ and different inverse temperatures $\beta$. 
An instantaneous $XX$ pair is either a $S$ or $S^\dagger$ operator, while $J_\text{ret}$ can connect different types of $X$ operators. The average number of kink-operators $X^{\alpha\beta}_\sigma X^{\beta\alpha}_\sigma$ (a subset of the retarded operators) is listed as $n_\text{kink}$.  
 The sign drops exponentially with $\beta$ and, at least in the temperature range considered, faster than exponentially with the average order of retarded  $XX$ pairs. This limits the simulations of the full model to high temperatures. 
In the following, we will therefore focus on the simplified model, with only instantaneous spin-flips and pair-hoppings.

\begin{figure}
	\includegraphics{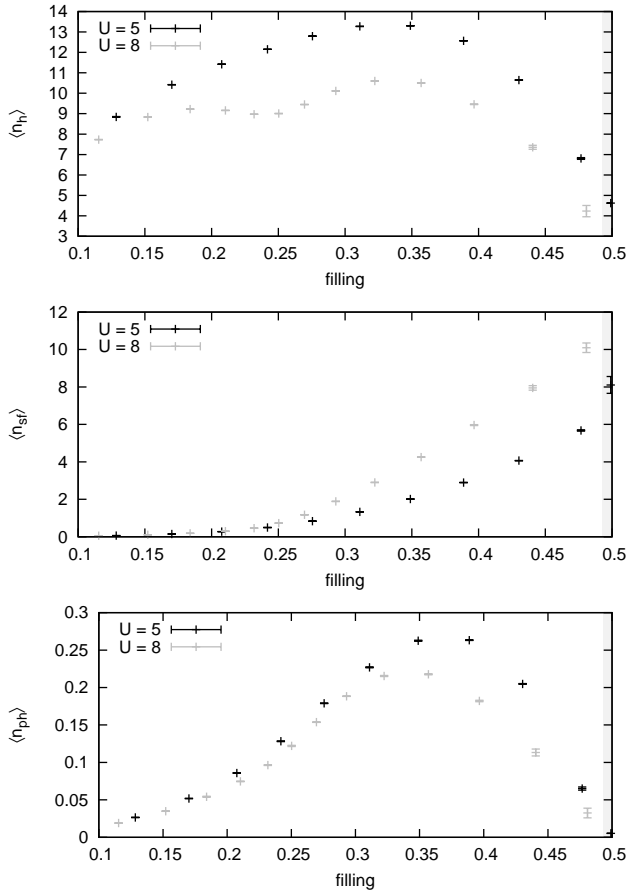}
	\caption{\label{fig_J_order} 
	Average perturbation order for $n_h$, $n_\text{sf}$, and $n_\text{ph}$ as a function of filling (per spin and orbital) for $U=5,8$ and $J=U/6$ ($\beta=50$).
}
\end{figure}

\begin{figure}
	\includegraphics{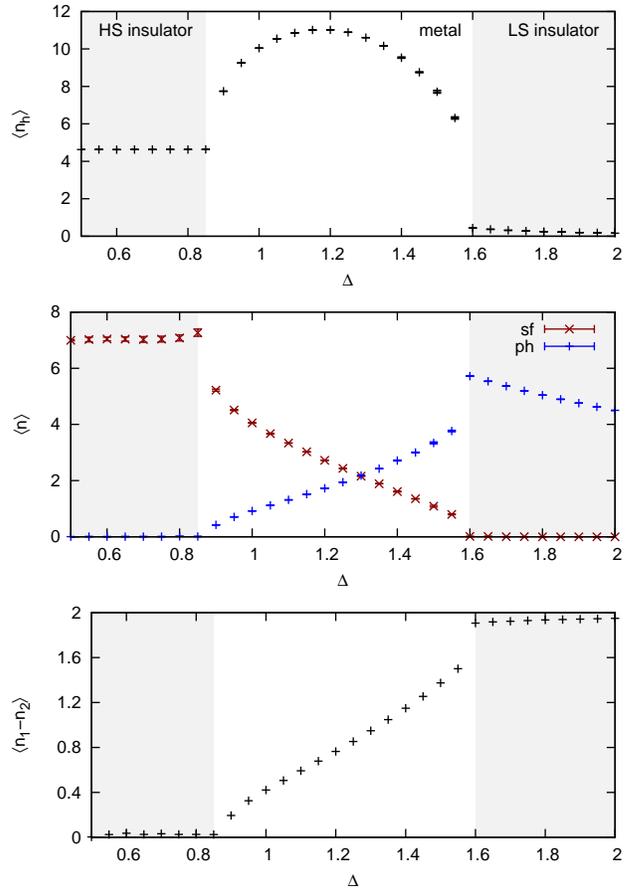}
	\caption{\label{fig_nmucf} 
	Top two panels: Average perturbation orders $n_h$, $n_\text{sf}$, and $n_\text{ph}$ as a function of the crystal field splitting $\Delta$ for $U=5$ and $J=U/6$ ($\beta=50$).  
	The lowest panel shows the orbital polarization $n_1-n_2$. 
}
\end{figure}

\section{Results}
\label{sec:results}

\subsection{Two-orbital model with static interactions}

We first consider the model with static interactions $U = 5$, $J = 5/6$  and $U=8$, $J=8/6$, respectively, and plot the average perturbation orders for the 
hybridization, spin-flip and pair-hopping terms as a function of filling (per orbital and spin) in Fig.~\ref{fig_J_order}. Near half-filling, and for the stronger interaction also near quarter-filling, 
the average hybridization order, and hence the kinetic energy, is suppressed. This suppression appears to be related to the spin-freezing phenomenon.\cite{Werner2008}
As was shown in previous studies,\cite{Medici2011, Hafermann2012} also the two-orbital model (both with and without spin-flip and pair-hopping terms) exhibits a bad metallic state with ``frozen" local moments in a certain filling range close to half-filling and quarter filling. The downturns near filling $\tfrac14$ and $\tfrac12$ in the hybridization order coincide with the onset of this spin-freezing regime (see Fig. S3 in Ref.~\onlinecite{Medici2011} and Fig. 12 in Ref.~\onlinecite{Hafermann2012}). 

While the pair-hopping order shows a similar down-turn, which is also explained by the appearance of $S=1$ moments, the spin-flip order increases systematically with filling and takes its largest value in the
half-filled Mott insulator. This is because for $J>0$, the Mott insulator is dominated by high-spin states with one electron in each orbital. While the spin-flip operators can act on some of these spin-triplet states, the pair-hopping operator cannot. The situation would be opposite for $J<0$.  

If the crystal field splitting $\Delta$ is increased in the half-filled high-spin Mott insulator, one either observes a first order transition to the low-spin insulator (at large $U$), or first a transition into a strongly correlated metallic state, followed by a second transition to the low-spin insulator.\cite{Werner2007crystal} For the latter case ($U=5$, $J=U/6$, $\beta=50$) we plot the different perturbation orders and the orbital polarization as a function of $\Delta$ in Fig.~\ref{fig_nmucf}. The hybridization order $n_h$ increases after the transition into the metal, since it is proportional to the kinetic energy, and becomes very low in the orbitally polarized low-spin insulator. The spin-flip order $n_\text{sf}$ remains approximately independent of $\Delta$ in the high-spin insulator, decreases continuously with increasing $\Delta$ in the metal, and then drops to very small values in the low-spin insulating phase. The average pair hopping order $n_\text{ph}$, which is very low in the high-spin insulator, grows with increasing orbital polarization in the metal phase, and then jumps to a maximum value on the insulating side of the metal-low-spin insulator phase boundary. While $n_\text{ph}$ is large in the low-spin insulating phase, it decreases with increasing $\Delta$, because the population of the higher orbital with two electrons becomes increasingly costly.

\subsection{Two-orbital model with dynamically screened $J$}

In this section, we present results for a two orbital model with static $U$ and dynamically screened $J$. To avoid the sign problem associated with retarded spin-flip or pair-hopping terms, we 
only treat the $J_z$ component dynamically ($J_z(\omega)=\mathcal{J}(\omega)$) and approximate the spin-flip and pair-hopping terms with a static $J_\text{scr}=\mathcal{J}(\omega=0)$. For the frequency dependence we adopt the single boson model (\ref{singleboson}). Since we will be interested in particular also in negative $J_\text{scr}$, our calculations may be viewed as simple model calculations for the alkali-doped fullerides. In these materials, the effectively negative $J_\text{scr}$ favors a low-spin state containing an intraorbital electron pair, which has been argued to drive the $s$-wave superconductivity next to the Mott insulating phase.\cite{Capone_2002,RevModPhys.81.943,nomura_C60_paper} Here, we will not study superconductivity (for a brief discussion of technical aspects related to simulations in the superconducting phase, see Appendix~\ref{app:super}), but the spin state transitions which occur as $J_\text{scr}$ is varied.

\begin{figure}
	\includegraphics[width=8.5cm]{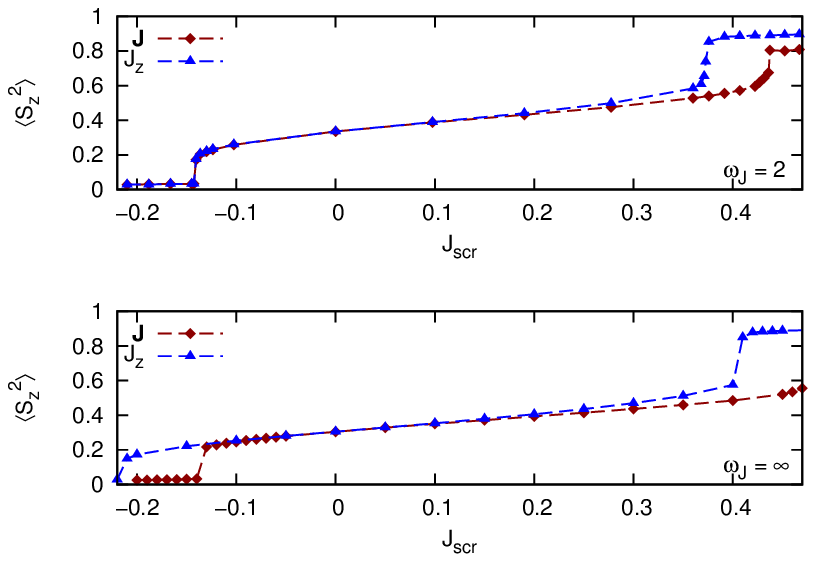}
	\includegraphics[width=8.5cm]{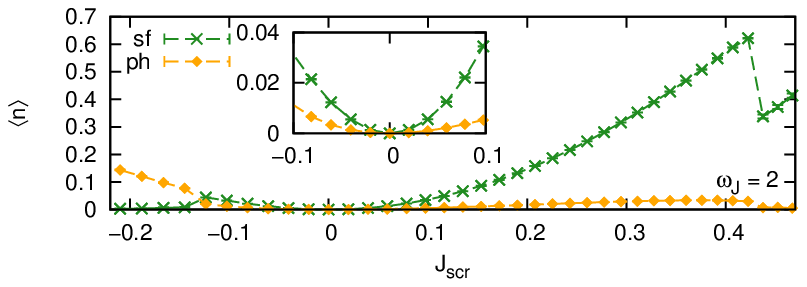}
	\includegraphics[width=8.5cm]{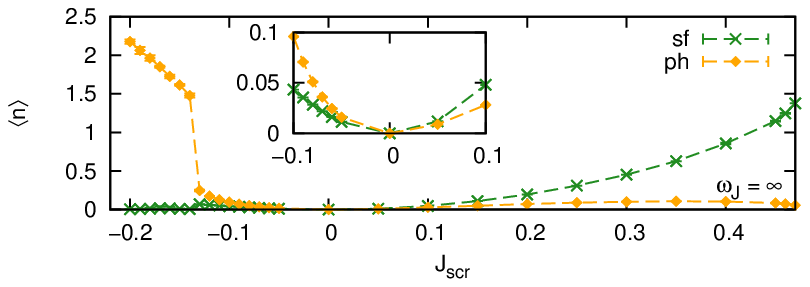}
	\caption{\label{fig_szcomp} 
	First two panels: $\langle S_z^2\rangle$ as a function of $J_\text{scr}$ for $U = 4$, $J_\text{bare}=1$, $\omega_J = 2$ (first panel) and $\omega_J=\infty$ (second panel), $\mu = \mu_{1/2}$ and $\beta = 25$. Both results for the simplified model with static spin-flip and pair-hopping terms (``$J$") and for the density density approximation (``$J_z$") are shown. Third and fourth panel: average perturbation orders for the spin-flip and pair-hopping terms in the simulation with $\omega_J=2$ (third panel) and $\omega_J=\infty$ (fourth panel).  
}
\end{figure}

The top panel in Fig.~\ref{fig_szcomp} plots $\langle S_z^2 \rangle$ for a half-filled model with $U=4$, unscreened $J_\text{bare}=1$, screening frequency $\omega_J=2$ and several values of $J_\text{scr}$ in the range $-0.2<J_\text{scr}<0.5$. The curve labeled by `$J$' shows the results of simulations with (static) spin-flip and pair-hopping terms. For sufficiently large $J_\text{scr}$, the half-filled system is in a high-spin insulating state and $\langle S_z^2\rangle \approx 1$. Around $J_\text{scr}=0.43$, a transition to a metallic phase occurs, and this phase is stable down to $J_\text{scr}\approx -0.13$. Only for screened values below $-0.13$ do we find a low-spin insulating phase. The curve labeled by `$J_z$' shows the results obtained from the density-density approximation (no spin-flip and pair-hopping terms). Consistent with Fig.~\ref{fig_gfcomp}, the stability range of the high-spin insulator is enhanced in the $J_z$ approximation. On the other hand, the transition to the low-spin insulator occurs at almost the same $J_\text{scr}$ as in the calculation with spin-flips/pair-hoppings. This is because in the range $-0.2 \lesssim J_\text{scr} \lesssim 0.2$ the average perturbation order for pair-hoppings and spin-flips is very low, as shown in the third panel.  
Note that in the simplified model, $J_\text{scr}$ is used for the spin-flip and pair-hopping terms, while the density-density part is dynamical (i.e. approaches $J_\text{bare}$ at high frequencies). This explains why for small negative $J_\text{scr}$, the spin-flip order 
is slightly larger 
than the pair-hopping order.
It also explains the drop in the perturbation order across the transition from the metal to the high-spin insulator: 
The simplified model 
breaks the spin SU(2) symmetry and hence the three-fold degeneracy of the high-spin states. 
In the high-spin insulating phase, the weights of the states 
$\ket{ \uparrow \uparrow }$ and $\ket{ \downarrow \downarrow }$ dominate.
Consequently, across the transition into the high-spin insulating state, the weights of the states 
$\ket{ \uparrow \downarrow }$ and $\ket{ \downarrow \uparrow }$ decrease. 
Since the spin-flip operators act on these states, the spin-flip perturbation order decreases 
as one enters into the high-spin insulating phase.

For comparison, we also show the results for the static limit ($\omega_J=\infty$) in the 
second and fourth panel. 
As expected, the high-spin insulator becomes less stable, since the transition occurs at $J_\text{scr}<J_\text{bare}$. 
The transition to the low-spin insulator 
exhibits a rather large 
difference in the critical $J_\text{scr}$ between the `$J$' and `$J_z$' approximations. 
This can be ascribed to the fact that the average perturbation orders for the pair-hopping and spin-flip interactions becomes larger
 compared to the $\omega_J=2$ case. In the $J_z$ approximation, the metallic solution 
 is stabilized in the vicinity of the transition to the low-spin insulator.  
 This result is qualitatively similar to the behavior of the Holstein-Hubbard model near the transition to the bipolaronic insulator, where a large screening frequency stabilizes the metallic phase (see e.~g. Fig.~2 in Ref.~\onlinecite{Werner_2010}). One can explain this correlation effect by translating the frequency-dependent density-density interaction into an effective bandwidth reduction.~\cite{Casula_2012_effmodel} 
 
\begin{figure}
	\includegraphics[width=8.5cm]{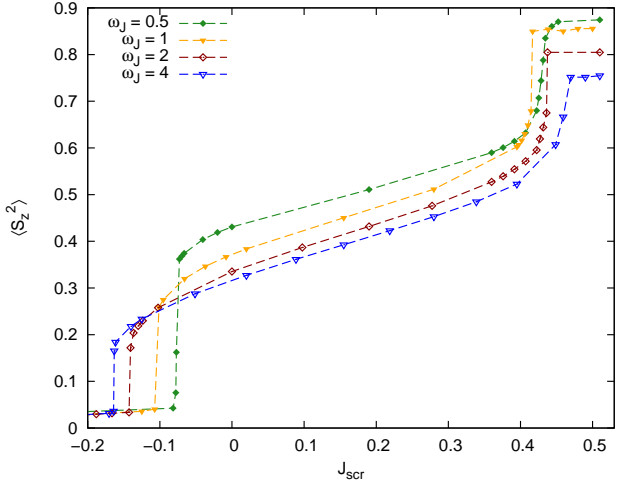}
	\caption{\label{fig_ss} 
	$\langle S_z^2\rangle$ as a function of $J_\text{scr}$ for $U = 4$, $J_\text{bare}=1$, $\omega_J = 2$, $\mu = \mu_{1/2}$, $\beta = 25$ and different screening frequencies $\omega_J$. 
}
\end{figure} 
 
In the calculation with spin-flips and pair-hoppings, the effect of $\omega_J$ on the critical $J_\text{scr}$ for the transition to the low-spin insulator is much smaller 
(the transition occurs at almost the same $J_\text{scr}$ for $\omega_J=\infty$ and $\omega_J=2$).  
This seemingly small effect is due to a compensation between the above-described stabilization mechanism for the metallic solution and 
the increase of the pair-hopping perturbation orders, which favor the insulating solution.

\begin{figure}
	\includegraphics[width=8.5cm]{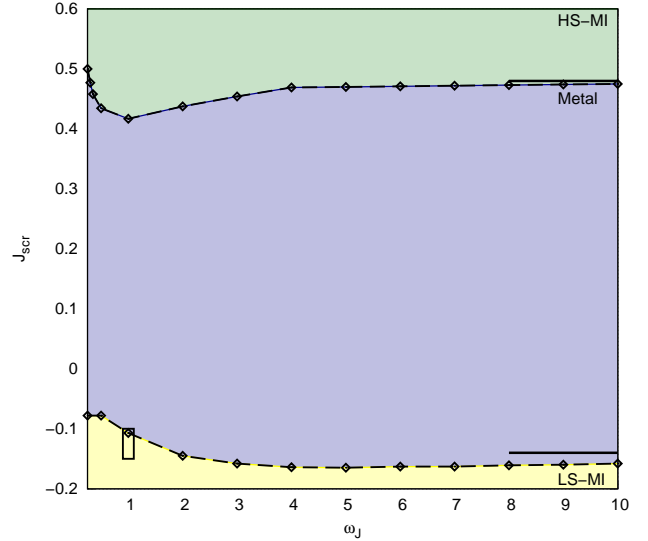}
	\caption{\label{fig_phase_diag} 
	Phase diagram in the space of $J_\text{scr}$ and $\omega_J$ for $U = 4$, $J_\text{bare}=1$, $\mu = \mu_{1/2}$ and $\beta = 25$. The rectangle at $\omega_J=1$ indicates the  parameter values for ``fullerene compounds" (see text). 
}
\end{figure}

For the calculation with spin-flip/pair-hopping terms, we plot $\langle S_z^2\rangle$ for different screening frequencies $\omega_J$ in Fig.~\ref{fig_ss}. With decreasing $\omega_J$ $(\lesssim 4)$, the transition to the low-spin insulator shifts to less negative values of $J_\text{scr}$, while the critical value for the transition to the high-spin insulator shows a non-uniform behavior. 
Also, in the limit of small $\omega_J$, the transition to the low-spin insulator is marked by a large jump in $\langle S_z^2\rangle$. 
The dependence of the phase boundaries on $\omega_J$ is plotted over a wider range of $\omega_J$ in Fig.~\ref{fig_phase_diag}. As $\omega_J$ is reduced from the large-frequency limit, both phase boundaries shift to smaller $J_\text{scr}$, because the increasing effect of the $J_\text{bare}=1$ on fast spin fluctuations stabilizes (destabilizes) the high-spin (low-spin) insulator. 
At smaller $\omega_J$, the band renormalization effect, which enhances the stability of the low-spin insulator, reverses this trend.

While a discussion of fullerene compounds based on results for a two-orbital model is dangerous, because the fullerides are three-orbital systems, it is still interesting to comment on the realistic values of $\omega_J$, $J_\text{scr}$, and $U$, if translated to the current set-up (with bandwidth $4$):  
$\omega_J \sim 1$, $J_\text{scr}$ between $-0.15$ and $-0.1$, and $U \sim 5$-$10$.\cite{PhysRevB.85.155452,nomura_C60_paper}
Our results in Fig.~\ref{fig_ss} and \ref{fig_phase_diag} suggest that the alkali-doped fullerides are located near the transition between the metal and 
low-spin insulator (see box in Fig.~\ref{fig_phase_diag}), in agreement with 
experiments.\cite{A15_CsC60nmat,Takabayashi20032009,fcc_CsC60,C60_exp_2015,PhysRevLett.104.256402,0295-5075-94-3-37007,PhysRevLett.112.066401}

The effect of a larger $U$ on the phase diagram of Fig.~\ref{fig_phase_diag} is to enhance the stability region of the high-spin insulator, and to a lesser extent also of the low-spin insulator. For example, for $U=6$, the transitions to the high-spin (low-spin) insulator occur at $J_\text{scr}= 0.13$ ($J_\text{scr}=-0.07$) for $\omega_J=\infty$, and at $J_\text{scr}=0.06$ ($J_\text{scr}=-0.03$) for $\omega_J=2$.

\section{Conclusions and Outlook}\label{sec:conclusions}

We have presented a double-expansion algorithm for multi-orbital impurity models which combines a hybridization expansion with a weak-coupling expansion for the spin-flip and pair-hopping terms. This algorithm is based on the economical segment representation and avoids computationally expensive matrix multiplications. By construction, it performs particularly well in the regime of weak Hund coupling. 

A main motivation for introducing this algorithm is the ability to treat dynamically screened $J(\omega)$. We have explained the Monte Carlo procedures for the sampling of retarded density-density, spin-flip and pair-hopping terms. In practice, however, the sign problem associated with retarded spin-flips and pair-hoppings forced us to consider a simplified model, in which only the $J_z$ component is treated as dynamical, while spin-flip and pair-hopping terms are approximated as static operators. This approximation allows sign-free simulations of the two-orbital model, and to explore the effect of a dynamically screened $J(\omega)$ in a set-up which recovers the rotationally invariant interaction in the limit of large screening frequency. 

Extending our formalism to models with more than two orbitals requires additional Monte Carlo moves which change the ``winding number" of configurations (Appendix~\ref{app:three_orb}). These updates will introduce negative weight configurations even in the case of the simplified model. Keeping track of the complex topology of the segment configurations for models with more than two orbitals becomes challenging. It may thus be simpler to pursue an alternative approach, which is not as efficient as the segment-based algorithm, but more flexible and easier to implement: As explained in Appendix~\ref{app:aux_method} one can rewrite the spin-flip and pair-hopping operators using auxiliary spin variables. In combination with the hybridization expansion algorithm in the matrix formulation and a weak-coupling expansion in the spin-flip and pair-hopping terms, this auxiliary field sampling will allow to generate configurations with an odd number of operators, and hence non-zero winding number. However, the physical configurations are in this approach always combined with fictitious ones, whose contributions to the Monte Carlo measurements average to zero, but exacerbate the sign problem.  

On the application side we have presented results for the spin state transitions in a two-orbital model with static $U$ and dynamically screened $J(\omega)$. These results show that our simplified model with static spin-flip and pair-hopping terms produces results which differ significantly from the density-density approximation near the transition to the high-spin Mott insulating state.
On the other hand, the transition to the low-spin insulating state occurs at small negative $J_\text{scr}$. Because of this, the perturbation orders for the spin-flip and pair-hopping terms are small, which results in almost identical transition points as in the density-density case. At least in our single-boson model, the low-spin insulating phase is stabilized significantly in the limit of low screening frequency $\omega_J$. Choosing realistic parameter values for fullerides, we find that within our two-orbital description, the system is on the verge of the transition from the metal to the low-spin insulating phase. 

Because of the relevance of multi-orbital models with overscreened $J$ for the physics of alkali-doped fullerides, an interesting future application will be the study of superconductivity in this model. In particular, it will be possible to investigate the effect of the frequency-dependent Hund's coupling on the properties of the superconducting state. To map out the stability regions of superconducting or other symmetry-broken  phases, it is sufficient to measure appropriate susceptibilities in the symmetric state, using the algorithm discussed in this paper. In order to enter the superconducting phase, additional Monte Carlo updates are needed. We briefly discuss this issue in  Appendix~\ref{app:super}. Testing the efficiency of the proposed method in the superconducting state is an important and interesting future problem.

\acknowledgements
	We thank S. Hoshino and L. Huang for helpful discussions. The simulations were run on the BEO04 cluster at the University of Fribourg, using a code based on ALPS.\cite{Albuquerque2006} This work was supported by SNF grant No.~200021-140648.

\appendix
\section{Three-orbital model}\label{app:three_orb}

For multi-orbital models with more than two orbitals, the updates described in Sec.~\ref{MCsampling} are not sufficient for an ergodic sampling. This is because configurations with non-zero ``winding number" contribute to the partition function, and these configurations cannot be generated by the insertion/removal of operator pairs $S S^\dagger$,  $S^\dagger S$ or $P P^\dagger$, $P^\dagger P$. We will focus the discussion here on the three-orbital case and the spin flip operators; the generalization to more orbitals and to the pair-hopping case is straightforward.  

\begin{figure}
	\includegraphics[width=8.5cm]{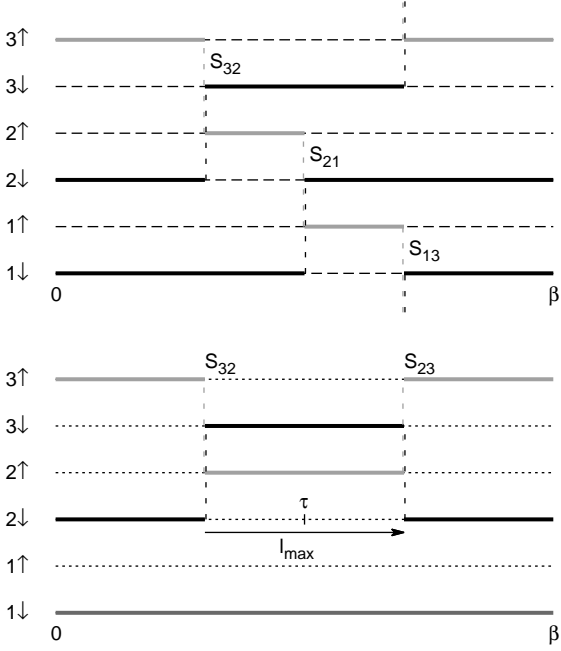}
	\caption{\label{fig:winding} 
	Top panel: Configuration with $n_S=3$ spin flips and nonzero winding number in the three-orbital model. Bottom panel: Configuration with $n_S=2$ spin-flips from which the configuration in the upper panel can be obtained using the procedure described in Appendix~\ref{app:three_orb}. 
}
\end{figure}

In the three orbital context, it is natural to work with the spin-flip operators $S_{\alpha\beta}=c^\dagger_{\alpha\downarrow} c_{\beta\downarrow} c^\dagger_{\beta\uparrow} c_{\alpha\uparrow}$ ($\alpha, \beta=1,2,3$ and $\alpha\ne \beta$). The number of these operators will be denoted by $n_S$.  An example of a $n_\text{S}=3$ configuration with operators 
$S_{32}$, $S_{21}$ and $S_{13}$
is illustrated in the top panel of Fig.~\ref{fig:winding}. 

To produce this configuration, we can start from a neighboring 
$S_{32}$, $S_{23}$ 
pair of spin flips as shown in the bottom panel. These operators define an interval of length $l_\text{max}$ in which we randomly choose the time $\tau$. We then propose to insert a spin flip 
$S_{21}$ 
at $\tau$, and simultaneously replace the  
$S_{23}$ 
operator by an 
$S_{13}$
operator. The proposal probability for this move is $p^\text{prop}(n_S\rightarrow n_S+1)=\frac{1}{n_S}\frac{d\tau}{l_\text{max}}$. (Of course, the configuration in orbital 
1 
must be compatible with this operator insertion and operator replacement, otherwise the move is rejected.) For the inverse move, we randomly select a spin-flip operator and propose to remove it by simultaneously changing the type of the spin-flip operator to the right. In this case, the proposal probability is $p^\text{prop}(n_S+1\rightarrow n_S)$, and the ratio of acceptance probabilities becomes
\begin{equation}
R(n_S\rightarrow n_S+1) = -\frac{n_S l_\text{max}J}{n_S+1}.
\end{equation}
Similarly, we could have proposed to insert a 
$S_{13}$
operator and to replace the 
$S_{23}$
by an 
$S_{21}$ (which is possible only if the \mbox{$1\!\uparrow$} state is occupied in orbital $1$). 
Note the minus sign in the acceptance ratio, which indicates that the sampling of the configurations with nonzero winding numbers introduces a sign problem. 

\section{An alternative way to perform the double expansion: ``auxiliary spin" method}
\label{app:aux_method}

Here, we propose an alternative method to treat spin-flip and pair-hopping interactions. 
For simplicity, let us consider static spin-flip and pair-hopping terms. 
In the scheme discussed in Sec.~\ref{sec_static} for the two-orbital model with static spin-flip and pair-hopping interactions, 
 the corresponding perturbation orders are always even: The $S$ operator is always paired with a $S^{\dagger}$ operator and the same 
 is true for $P$ and $P^{\dagger}$. 

 Alternatively, we can perform the simulation by introducing auxiliary spins,   
 similarly to the strategy employed in Ref.~\onlinecite{PhysRevB.80.155132} for the interaction-expansion impurity solver.\cite{note_nomura} 
 In this alternative method, we rewrite an ${\mathcal O}$ operator 
(${\mathcal O} = S, S^{\dagger}, P,$ and $P^{\dagger}$) as 
\begin{eqnarray}
 {\mathcal O} = \frac{1}{2}\sum_{s_a = \pm 1}  \tilde{\mathcal O}_{s_a},  
\end{eqnarray}
where 
\begin{eqnarray}
\tilde{\mathcal O}_{s_a}   =  {\mathcal O} + s_a \gamma I
\end{eqnarray}
with $s_a$ the auxiliary spin, $\gamma$ a positive real number, and $I$ the identity operator. 
We then expand the partition function in powers of the hybridization operators and the $\tfrac12 \tilde{\mathcal O}_{s_a} $ operators 
and perform the sum over the spin orientations $s_a =\pm 1$ in the Monte Carlo sampling. 

In this scheme, in addition to the insertion/removal of pairs of hybridization operators, we use the following updates: 
\begin{enumerate}
\item insertion and removal of $\tfrac12 \tilde{S}_{s_a}(\tau_s)$, 
\item insertion and removal of $\tfrac12 \tilde{S}^{\dagger}_{s_a}(\tau_s)$, 
\item insertion and removal of $\tfrac12 \tilde{P}_{s_a}(\tau_p)$, 
\item insertion and removal of $\tfrac12 \tilde{P}^{\dagger}_{s_a}(\tau_p)$. 
\end{enumerate}
Note that, in this method, the perturbation order can be odd because of the $s_a \gamma I$ term. 
In evaluating the configuration weights, if we employ the segment formalism, we have to 
sum up the weights for $2^N$ configurations with $N$ being the total perturbation order of the $\tfrac12 \tilde{\mathcal O}_{s_a} $ operators. 
This is because at each $\tilde{\mathcal O}_{s_a}$ vertex, we have the choice between $ {\mathcal O}$ and $s_a \gamma I$ operators. 
It is easy to show that many of these $2^N$ configurations have zero weight.  
Therefore, in practice, we can reduce the number of configurations in the calculation of the weight. 
Another possibility is to employ the matrix formalism.\cite{Werner_2006_Matrix} 
In this case, each of the $\tfrac12 \tilde{\mathcal O}_{s_a} $ operators can be represented by a single matrix,  
and the summation of the weights over $2^N$ configurations can be avoided. 
In practice, if we use a block-diagonalized matrix representation and local conserved quantities,~\cite{Haule_2007,PhysRevB.86.155158} 
we only need to apply the $s_a \gamma I$ operator to block-diagonalized matrices which contain nonzero matrix elements of the operator ${\mathcal O}$.
A larger $\gamma$ improves the acceptance ratio of the above-mentioned updates; however, 
a larger $\gamma$ also produces a larger number of samples with negative weight.
Therefore, one has to find the optimal value for $\gamma$, which will depend on e.g. the size of $J_{\rm scr}$.

In the presence of retarded density-density interactions, the configuration weight also contains a bosonic factor (Sec.~\ref{sec_ret_dens}). In the auxiliary spin algorithm, we evaluate this bosonic factor with the operators  $\mathcal{O}$ instead of $\tilde{\mathcal{O}}_{s_a}$. This procedure produces unphysical weights for configurations involving $s_a\gamma I$ operators due to an inconsistency in evaluating the local trace and the bosonic factor. However, these contributions will average to zero in the Monte Carlo sampling, and thus we still get correct thermal averages for the physical quantities.  
Because of the ``fictitious" weights involving $s_a\gamma I$ operators, it is clear that the auxiliary spin method will have a more severe sign problem, and thus be less efficient than the 
scheme described in Sec.~\ref{MCsampling}. 
However, it circumvents the algorithmic complexity originating from the increase in the number of orbitals, as described in Appendix~\ref{app:three_orb}.

\begin{figure}[t]
	\includegraphics[width=8.5cm]{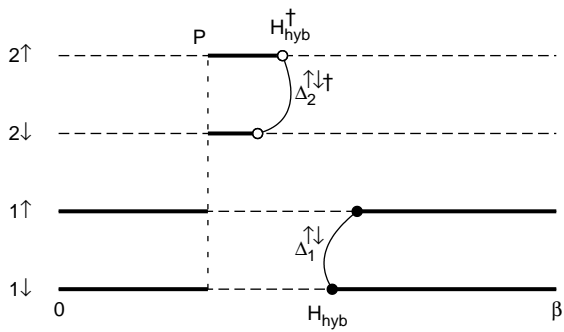}
	\caption{Illustration of a configuration with one pair-hopping operator and two anomalous hybridization functions $\Delta_\alpha^{\uparrow\downarrow}$ which contributes to the partition function of a two-orbital model with intra-orbital singlet pairing. This configuration can be obtained from the one with two full lines in orbital $\alpha=1$ by the procedure described in Appendix~\ref{app:super}. 
}
\label{fig:supercond} 
\end{figure}

\section{Simulations in the superconducting state}\label{app:super}

In a simulation of a symmetry-broken phase, we sometimes need special updates in addition to the updates used in the simulation of the normal phase, as was pointed out in Ref.~\onlinecite{PhysRevB.89.165113}. For example, let us consider the case of $s$-wave superconductivity in a two-orbital model with a negative static $J_{\rm scr}$. A negative $J_{\rm scr}$ favors intraorbital singlet pairing.\cite{Capone_2002,RevModPhys.81.943,Koga2015,nomura_C60_paper}
To sample this superconducting state, one will need additional updates
such as the insertion and removal of a $P$ [$P^{\dagger}$] operator and two anomalous hybridization functions (see Fig.~\ref{fig:supercond}). These anomalous hybridization functions correspond to two ${\mathcal H}_{\rm hyb}$ operators with flavor 
$(1,\uparrow)$ and $(1,\downarrow)$ [$(2,\uparrow)$ and $(2,\downarrow)$] 
and two  ${\mathcal H}^{\dagger}_{\rm hyb}$ operators with flavor 
$(2,\uparrow)$ and $(2,\downarrow)$ [$(1,\uparrow)$ and $(1,\downarrow)$]. 
Alternatively, one could think of an update in which one instantaneous $P$ [$P^{\dagger}$] operator is replaced by 
two ${\mathcal H}_{\rm hyb}$ operators with flavor 
$(2,\uparrow)$ and $(2,\downarrow)$ [$(1,\uparrow)$ and $(1,\downarrow)$] 
and two ${\mathcal H}^{\dagger}_{\rm hyb}$ operators with flavor 
$(1,\uparrow)$ and $(1,\downarrow)$ [$(2,\uparrow)$ and $(2,\downarrow)$]. 

\bibliography{hw}

\begin{thebibliography}{58}
\bibitem[{not({\natexlab{b}})}]{note_nomura}
\bibinfo{note}{We note that an improved method to treat the spin-flip and
  pair-hopping terms in the interaction-expansion impurity solver has been
  proposed in Ref.~\onlinecite{PhysRevB.89.195146}.}
\end{thebibliography}

\end{document}